\documentclass[12pt]{iopart}
\usepackage{enumerate}
\usepackage{subfigure}
\usepackage{iopams}
\usepackage{amstext}
\usepackage{psfrag}
\usepackage{graphicx}
\usepackage{harvard}

\newcommand{\be}{\begin{equation}}
\newcommand{\ee}{\end{equation}}
\newcommand{\bel}[1]{\begin{equation}\label{#1}}
\newcommand{\bea}{\begin{eqnarray}}
\newcommand{\eea}{\end{eqnarray}}
\newcommand{\ba}{\begin{array}}
\newcommand{\ea}{\end{array}}

\newcommand{\eel}{\end{equation}}

\newcommand{\bra}[1]{\mbox{$\langle \, {#1}\, |$}}
\newcommand{\ket}[1]{\mbox{$| \, {#1}\, \rangle$}}
\newcommand{\exval}[1]{\mbox{$\langle \, {#1}\, \rangle$}}

\def\bbbr{{\rm I\!R}} 
\def\bbbn{{\rm I\!N}}

\def\bbbz{{\mathchoice {\hbox{$\sf\textstyle Z\kern-0.4em Z$}}
{\hbox{$\sf\textstyle Z\kern-0.4em Z$}}
{\hbox{$\sf\scriptstyle Z\kern-0.3em Z$}}
{\hbox{$\sf\scriptscriptstyle Z\kern-0.2em Z$}}}}

\eqnobysec

\begin{document}

\review{Fluctuation theorems for stochastic dynamics}
\author{R J Harris$^1$ and G M Sch{\"u}tz$^2$}
\address{$^1$ Fachrichtung Theoretische Physik, Universit\"at des Saarlandes,
 66041 Saarbr\"ucken, Germany}
\address{$^2$ Institut f\"ur Festk\"orperforschung, Forschungszentrum 
J\"ulich, 52425 J\"ulich, Germany}
\eads{\mailto{harris@lusi.uni-sb.de}, \mailto{g.schuetz@fz-juelich.de}}

\begin{abstract}

Fluctuation theorems make use of time reversal to make predictions
about entropy production in many-body systems far from thermal
equilibrium.  Here we review the wide variety of distinct, but
interconnected, relations that have been derived and investigated
theoretically and experimentally.  Significantly, we demonstrate, in
the context of Markovian stochastic dynamics, how these different
fluctuation theorems arise from a simple fundamental time-reversal
symmetry of a certain class of observables. Appealing to the notion of
Gibbs entropy allows for a microscopic definition of entropy
production in terms of these observables.  We work with the master
equation approach, which leads to a mathematically straightforward
proof and provides direct insight into the probabilistic meaning of
the quantities involved. Finally, we point to some experiments that
elucidate the practical significance of fluctuation relations.

\end{abstract}


\submitto{Journal of Statistical Mechanics: Theory and Experiment}

\maketitle

\pagestyle{plain}

\noindent\hrulefill

\tableofcontents

\noindent\hrulefill

\section{Introduction}
\label{s:intro}

Nobody (or so we believe) has ever observed that somewhere in their
office the gas molecules in a cubic centimetre of air suddenly
dispersed to leave empty the entire volume they had previously
occupied.  In fact, the time reversibility of Newtonian dynamics
allows for such a dramatic event which simply requires the gas
molecules to follow time-reversed trajectories in phase-space
(compared to the familiar trajectories which lead to filling an
artificially prepared vacuum).  However, while possible in principle,
the corresponding probability of these time-reversed trajectories is
just too small to make any claim to the effect credible.  Nor would
we expect a spontaneous cooling down of some macroscopic object by a
few degrees, followed by an equally spontaneous return to room
temperature an hour later.  Even though we could perhaps imagine this
to happen in a tiny system of just a few molecules for a very short
spell of time, we would be greatly surprised if such an abnormal state
of matter persisted for a long period.   Nevertheless, phenomena like
these are what the fluctuation theorems that we review here are all
about, at least in their mathematically pure form.

So why do we care about rare events which may occur on atomic length
and time scales but which quickly become exceedingly improbable as we
turn to macroscopic observation scales? The first and simplest answer is
that small observation scales are becoming accessible to experiments
and so probabilities that have traditionally been irrelevant begin to
play a role for our understanding of microscopic physics.

To give a second and deeper answer, we note that in thermal
equilibrium such rare events actually do play a central role, not
directly in experiments, but as a tool for the theoretical foundation
of equilibrium thermodynamics.  Explaining this requires a reminder of
basic thermodynamics, of what probabilists call large deviation
theory.  In the microcanonical ensemble one considers the number of
microstates of an \emph{isolated} system at some fixed value of internal
energy $U$, volume $V$ and other extensive conserved properties. The
logarithm of this number of microstates is the entropy $S$ (we set the
Boltzmann constant $k_B=1$) and by inverting this function one obtains
the internal energy $U(S,V,\dots)$. From this so-called thermodynamic
potential all other thermodynamic quantities (temperature, pressure,
heat capacity and so on) can be computed as a function of the
extensive quantities. More frequently, however, we are faced with the
situation of an \emph{open} system in exchange with a heat bath at
temperature $T$. This requires a description of the system in terms of
a different thermodynamic potential, the (Gibbs) free energy $F$. This
is obtained from $U$ by considering the probability that in a small
test volume of a microcanonical system (inside which the energy
fluctuates around the mean $U/V$) the local energy deviates strongly
(by a macroscopic amount) from its mean in that volume. This 
probability, exponentially small in the volume $V$, is the
subject of large deviation theory which underlies the passing from one
thermodynamic ensemble to another, with well-known everyday
consequences, relevant for the functioning of refrigerators or
automobile engines. The mathematical tool required for switching
between equivalent thermodynamic potentials is the
Legendre transformation which relates extensive quantities
to their nonextensive conjugate variables.

Unfortunately, such a powerful large deviation theory, based on an
explicit assumption of the form of the entropy, is not known far
from thermal equilibrium and can hardly be expected to exist in such
generality. Even for specific non-equilibrium systems little can
usually be said about the probability distributions, let alone about
their tails which are relevant for rare events. Therefore it came as a bit
of a surprise that non-equilibrium fluctuations theorems (FTs for
short) give at least some insight of very general nature. This is
encouraging and another reason for the recent strong interest in
FTs, even if their experimental accessibility and everyday
significance is comparatively low at the present early stage of the
development.

The key ingredient underlying all FTs is time reversal,
well-understood in thermal equilibrium but not so well-explored in
non-equilibrium systems. Time reversal is not sufficient to determine
the form of the large deviation function, but, in the context of FTs,
it is shown to provide information about the rate of entropy
production and thus sheds some light on non-equilibrium probability
distributions.  Historically, an explicit fluctuation theorem first
arose in the context of the simulation of sheared fluids by Evans,
Cohen and Morriss \cite{Evans93}. Shortly afterwards Gallavotti and
Cohen proved rigorously a closely related FT for deterministic
dynamics \cite{Gallavotti95}. Some initial scepticism about the
usefulness of such theorems was gradually
overcome. \citeasnoun{Jarzynski97} showed how to relate non-equilibrium
properties to equilibrium quantities.  \citeasnoun{Kurchan98} set up a
framework for deriving FTs for single-particle stochastic Langevin
dynamics and this was followed by an extension to fairly general
Markov processes by \citeasnoun{Lebowitz99}. Eventually (from the year
2000 or so onwards) a vast body of literature both on deterministic
and stochastic dynamics emanated from these ideas.  FTs for quantum
systems have also been considered, see e.g., \citeasnoun{Kurchan00},
\citeasnoun{Esposito06}.  From an experimental perspective, recent
years have seen an increasing number of results supporting the
validity of FTs \citeaffixed{Ritort03}{see e.g., the review by}.

At this point it is interesting to note that, in several parallel
developments, rare events in non-equilibrium many-body systems have
become a focus of intense research. They play a central role in
the study of extreme events, such as climate extremes, earthquakes,
and other potentially catastrophic phenomena \cite{Albeverio06}. An
important open question is whether extreme events occurring in vastly
different systems have any common dynamical and statistical
features. On a purely theoretical level, the tails of non-equilibrium
probability distributions have been examined with great success in the
framework of driven diffusive systems---simple stochastic lattice-gas
models of non-equilibrium interacting particle systems. Universal
properties of the distribution of the particle current have been
uncovered \cite{Praehofer02}, large deviation functions have been
computed explicitly for special models, revealing the correlated
structure of these non-equilibrium steady states \cite{Derrida05b,Derrida07b}, and
in a hydrodynamic approach, making explicit use of time reversal, a
fairly general large deviation theory has been developed \cite{Bertini06}.

The wealth of explicit exact results as well as the possibility of
numerical simulation make driven diffusive systems a fertile testing
ground for fluctuation theorems. Moreover, the mathematical framework
of the master equation for the probability distribution allows for
quite straightforward derivations of the multitude of known
fluctuation theorems from a single time-reversal relation. This relation
involves an observable which is intimately related to entropy production.

In our theoretical description we choose, mainly for these reasons, the 
master equation approach for the description of stochastic many-body 
dynamics and for illustration we occasionally refer to driven diffusive 
systems as examples. We wish to stress, however, that our review covers much more ground than lattice gases.
We do not require systems to have a conserved particle number and when
we speak about currents we do not generally refer to particle
currents. As we shall see, the currents under investigation
are related to entropy production in some way or other and the various
FTs provide different kinds of information about entropy production,
depending on the initial state and precise nature of the process.

In an experiment or in numerical simulation
the system under investigation is prepared at time
$t=0$ by some mechanism (that does not directly enter the theoretical
description) in the desired initial state. Then the stochastic dynamics
for which an FT is being investigated is run for some time $t$.
In general, the stochastic dynamics are determined by some experimental
protocol, i.e., a time-dependent tuning of transition rates of the 
stochastic
process. While
running the experiment the quantity of interest is measured and usually
also the final state of the process (at time $t$) is recorded.
For good statistics the experiment may have to be repeated very many
times. In general
it is \emph{not} assumed that the initial distribution or the final
state are in equilibrium with respect to the stochastic process under
investigation. In general we do not even require
stationarity. Equilibrium conditions or stationarity are
required only for some special FTs, as discussed below.
The quantities that are measured are usually
particle displacements in a non-equilibrium situation which together
with the applied forces provide information about the work performed
on the system, or currents
e.g. in electrical systems, from which again work and entropy-like
quantities can be computed and compared with predictions from FTs.

In order to go all the way from time reversal to actual experiments we
first introduce the mathematical framework for treating the stochastic
dynamics of many-particle systems (section~\ref{s:formalism}). The
distinctive feature of our approach is the explicit inclusion of the
measurement process in the generator for the stochastic dynamics. This
facilitates the derivation in section~\ref{s:key} of a fundamental FT
from which (directly or by straightforward generalizations) other 
specific FTs which appear in the literature follow. This derivation is
preceded by a thorough discussion of time reversal.  In the following
sections we review a number of FTs that have been put forward, valid
for finite time intervals (section~\ref{s:finite}) or in a somewhat
simplified form in the long-time limit in the case of constant rates
or time-periodic protocols (section~\ref{s:asymp}). In
section~\ref{s:exp} we present a brief discussion of some experiments
that demonstrate validity and usefulness of FTs.  We make no attempt
whatsoever at providing a complete review of all relevant
experiments. We conclude the review by a short (and again highly
selective) outlook and a short appendix that illustrates in a simple
concrete model how FTs work.

For some other reviews emphasizing different aspects of FTs we refer to 
\citeasnoun{Evans02b}, \citeasnoun{Ritort03}, \citeasnoun{Maes03b}, 
\citeasnoun{Kurchan05} and \citeasnoun{Bustamante05}.  In a broader
context, there is also good coverage in some lecture notes on
non-equilibrium statistical mechanics, notably the expositions by 
\citeasnoun{Gaspard06} (focusing on Hamiltonian dynamics) and 
\citeasnoun{Maes07} (lattice gases).

\section{Stochastic many-particle systems}
\label{s:formalism}

The presentation and discussion of fluctuation theorems below
follows widely used standard notation and the reader interested only
in the theorems and their application may skip reading this section.
Some derivations, however, require setting up a suitable mathematical 
framework for treating stochastic dynamics. Since
we are concerned with interacting many-body systems we employ
the powerful master equation approach in the quantum Hamiltonian formalism, 
outlined in this section. This approach has a long history going
back at least to \citeasnoun{Glauber63}. We point out that the term
``quantum Hamiltonian formalism'' is really no more than a fanciful expression
for writing a set of linear first-order differential equations in matrix
form, using quantum mechanical bra/ket notation for vectors and a 
clever (and natural) choice of basis. However, doing so not only simplifies 
notation (once one gets used to it), but uncovers a very useful 
relationship to many-body quantum systems
in various specific cases of wide interest. For more details we refer
to two reviews that highlight applications and connections to
field theory \cite{Mattis98} and integrable quantum systems \cite{Schutz01}.

\subsection{Master equation for interacting particle systems}

\subsubsection{State space}
\label{sss:state}

This review is concerned with Markov processes that describe
interacting particle systems whose microstates at a given time
$t$ are labelled $\sigma(t)$. As examples for which the
fluctuation theorems to be discussed apply, we consider lattice 
models with a local state space $V$. That means that
we can describe the microstates $\sigma$ in terms of local
state variables $\sigma_{(i)} \in V$ where $i \in \Lambda$ denotes
a site of the lattice $\Lambda$. Both $V$ and $\Lambda$
may be infinite sets. We shall often refer to a specific microstate
as a configuration. A statistical ensemble defined by some
probability measure on the full state space
$\mathbb{V} = V^{|\Lambda|}$ is referred to as the state of the 
system.\footnote{Notice that in this nomenclature a state
characterized by a Dirac measure concentrated on a configuration
$\sigma$ is the same as the configuration $\sigma$ itself. 
Hence sometimes we refer also to configurations as states.}
In most of the exposition we consider discrete state space.

A particular history of the process over the time interval $[0,t]$
is completely specified by $\{\sigma\} \equiv \{ \sigma(\tau),0 \leq
\tau \leq t \}$.  Alternatively (as often in the literature), this information
is encoded by a chronologically-ordered list of the configurations the system
occupies $\sigma_0, \sigma_1, \sigma_2, \ldots, \sigma_n$ together
with the times of the transitions between them $\tau_1,\tau_2, \ldots,
\tau_n$.  Here $n+1$ is the total number of configurations visited
which is a history-dependent number. This notation is particularly
useful for situations where the transition times themselves are
irrelevant. The set $\{\sigma\}$ is sometimes also called a 
(stochastic) trajectory of the process. Time $t>0$ is continuous. 

For illustration we mention lattice-gas models where $\sigma_{(i)}$ is the particle 
number at site $i$ of the lattice. For particle systems with several
distinct species of particles the quantity $\sigma_{(i)}$ is a vector
with components giving the occupation number for each type
of particle. In single-species exclusion processes each lattice
site is occupied by at most one particle (exclusion principle) and
$V = \{0,1\}$. In the absence of any exclusion one has 
$\sigma_{(i)} =0,1,2,\dots$, i.e., $V = \bbbn$.  We stress, however, 
that $\sigma_{(i)}$ may denote any physical quantity or set of quantities. 
For example, spin systems may be described in this fashion, for Ising or Kawasaki
type models one has $V = \{-1,1\}$.

\subsubsection{Stochastic dynamics}

In the Langevin approach one defines a stochastic process in terms of 
a stochastic differential equation for the stochastic variables 
$\sigma(t)$ which contains both deterministic and noisy force terms.
The statistical properties of the noise are postulated on 
phenomenological grounds. For a given noise evolution,
the trajectory $\{\sigma\}$ is
uniquely determined by the initial state $\sigma_0 \equiv \sigma(0)$ and
expectation values are computed by taking appropriate
averages over histories with different realizations of the noise
and averages over initial conditions.
In contrast, we describe the stochastic dynamics through the master 
equation approach in which a deterministic evolution for the 
probability distribution is postulated. The physical forces, both
deterministic and random,
are encoded in transition rates $w_{\sigma', \sigma}(t)$ given by the 
transition probability $p_{\sigma', \sigma}(t)=w_{\sigma', \sigma}(t)dt$ 
from $\sigma \to \sigma'$ in the infinitesimal time interval $[t, t+dt]$.
Transitions therefore occur after random times which are distributed according to the
time-integrated exponentials of the transition rates. All transitions
are statistically independent from the history of the process that
led to the current configuration, except for the explicit time-dependence
of the rates.

Such rates may be determined from transition state theory or other
microscopic theories and therefore represent effective interactions.
This is not the concern of this review. We shall regard 
these rates as given parameters of the model,
assuming throughout a weak reversibility condition in the sense
that if $w_{\sigma', \sigma}(t)>0$ then also $w_{\sigma,\sigma'}(t)>0$. 
This excludes processes that freeze after
some relaxation period into a finite set of absorbing configurations.
Sometimes it will be convenient to consider all rates as parameterized
by a single external variable $l$.  A ``protocol'' of a process
with time-dependent rates is then specified
by giving $l(t)$.

\subsubsection{Master equation}

The memoryless Markovian nature of the stochastic dynamics 
is captured in the master equation
\bel{2-1}
\frac{\rmd}{\rmd t} P(\sigma;t) = 
\sum_{\stackrel{\sigma'\in \mathbb{V}}{\sigma'\neq\sigma}}
\left[w_{\sigma, \sigma'}(t)P(\sigma';t) -
w_{\sigma', \sigma}(t)P(\sigma;t)\right]
\ee
for the time evolution of the probability $P(\sigma;t)$ of finding the configuration 
$\sigma$ at time $t$.
Notice that, since we did not specify the state space $\mathbb{V}$, the
``probability'' may actually be a probability density. Nevertheless,
except where confusion may arise,
we shall refer to $P(\sigma;t)$ as probability as a shorthand for
probability, probability density, or probability measure.

Particularly for interacting particle systems, a more convenient way to 
write the master equation (\ref{2-1}) is provided by the quantum Hamiltonian 
formalism; for a detailed review, see
\citeasnoun{Schutz01}. The idea is to assign to each of the possible configurations 
$\sigma$ a vector $\ket{\sigma}$ which together 
with the transposed vectors  $\bra{\sigma}$ form an orthogonal basis of
a complex vector space and its dual.
Therefore the probability distribution can be written as probability vector
\bel{2-3}
| \, P(t)\, \rangle = \sum_{\sigma \in \mathbb{V}} P(\sigma;t) \, \ket{\sigma}.
\ee
With the scalar product $\bra{\sigma}\,\sigma'\,\rangle= \delta_{\sigma,\sigma'}$
one then recovers the probability by $P(\sigma;t)= \bra{\sigma} \, P(t)\, \rangle$. 
In a natural choice of basis a configuration $\sigma$ of the entire lattice 
system with $L$ sites is represented by a tensor state
$\ket{\sigma} =  \ket{\sigma_{(1)}}\otimes\dots\otimes\ket{\sigma_{(L)}}$.

In this formalism one rewrites the master equation in the form of
a Schr\"odinger equation in imaginary time,
\bel{2-4}
\frac{\rmd}{\rmd t} \, \ket{P(t)} = - 
H \, | \, P(t) \, \rangle ,
\ee
where the off-diagonal matrix elements of $H$ are the (negative) transition 
rates $w_{\sigma,\sigma'}(t)$ between configurations and the diagonal entries 
are the sum of all outgoing transition rates $w_{\sigma',\sigma}(t)$ from 
configuration $\sigma$. A single transition $\sigma \to \sigma'$ is 
therefore generated by the matrix
\bel{2-4a}
H(\sigma',\sigma) = w_{\sigma',\sigma} (\, \ket{\sigma}\bra{\sigma} \,
- \, \ket{\sigma'}\bra{\sigma}\, ) \equiv D(\sigma',\sigma) - E(\sigma',\sigma).
\ee
Here the off-diagonal $E$ turns configuration $\sigma$ into $\sigma'$
while the diagonal matrix $D$ takes care of probability conservation.
The time-dependence has been suppressed in this expression.

The master equation is linear and therefore has a formally simple
solution. The probability vector at time $t_0 + t$ is given in 
terms of an initial state at time $t_0$ by
\bel{2-5}
| \, P(t_0+t) \, \rangle = T \{ \rme^{-\int_{t_0}^{t_0+t}
H(\tau) \, \rmd \tau} \} \, | \, P(t_0) \, \rangle.
\ee
Note that, for time-independent rates, the time-ordered exponential reduces to a
simple exponential. The conditional transition probability
\bel{2-5a}
P(\sigma';t|\sigma;0) = \bra{\sigma'} \, T \{ \rme^{-\int_0^t
H(\tau) \, \rmd \tau} \} \, \ket{\sigma}
\ee
is the sum of the normalized statistical weights of all trajectories from a
configuration $\sigma$ to a configuration $\sigma'$ for a time
interval of length $t$.

In order to compute expectation values, we introduce the summation
vector $\bra{s} = \sum_{\sigma \in \mathbb{V}} \bra{\sigma}$. In the
scalar product with the probability vector it performs the
average over all possible final states of the stochastic time evolution,
i.e., $\bra{s} \, P(t) \, \rangle=1$. 
The expectation value $\exval{A(t)} = \sum_{\sigma \in \mathbb{V}} 
A(\sigma)P(\sigma;t)$ of some function $A(\sigma)$
is then given by the formula $\exval{A(t)}=\bra{s}\, A \,\ket{P(t)}$.
Here $A=\sum_\sigma A(\sigma)\, \ket{\sigma}\bra{\sigma}$ is a diagonal matrix 
which has  $A(\sigma)$ as diagonal
elements. The angular brackets denote averages both over histories
and the initial distribution. Of course, if the initial distribution
is concentrated on a particular configuration, the brackets
reduce to averages over histories.

\subsubsection{Stationarity, ergodicity, reversibility}
\label{sss:ser}

In this subsection we focus mainly on processes with time-independent
transition rates.
By construction one has
$\bra{s} \, H = 0$ which expresses conservation of probability. This
relation also shows that there is at least one right eigenvector
$\ket{P^\ast}$ with vanishing eigenvalue. For time-independent rates
this is a stationary
probability vector which is unique in a ergodic system. 
Two-time expectation values $\exval{A_2(t_2) A_1(t_1)}^\ast$
evaluated in a stationary distribution
depend only on the time difference.

Since in general
$H$ is not symmetric, there is no simple relationship between left and
right eigenvectors of $H$. In stochastic dynamics that are ergodic
and satisfy
the detailed balance condition
\bel{2-6}
w_{\sigma, \sigma'}P^\ast(\sigma') =
w_{\sigma', \sigma}P^\ast(\sigma)
\ee
one has for the transposed evolution operator
\bel{2-7}
P^\ast H^T (P^\ast)^{-1} = H 
\ee
where $P^\ast$ is the diagonal matrix with $P^\ast(\sigma)$ as diagonal
elements. 
With this property one can easily prove reversibility of the
process which means that any stationary multi-time correlation function
with $t_{i+1}\geq t_i \geq 0$ satisfies
\bel{2-7a}
\exval{A_n(t_n)\dots A_2(t_2) A_1(t_1)}^\ast = 
\exval{A_1(t-t_1) A_2(t-t_{2}) \dots A_n(t-t_n)}^\ast
\ee
for any $t\geq t_n$.
We refer to stochastic particle systems satisfying detailed balance as 
equilibrium systems. An equilibrium system has a purely real relaxation spectrum
and can always be mapped using (\ref{2-7}) to a quantum
system with real symmetric Hamiltonian $H$ and ground state vector
related to $\ket{P^\ast}$. This is the origin
of the term quantum Hamiltonian formalism.

One can obtain information about the late-time behaviour of the
process, in particular, how it approaches stationarity, 
by inserting a complete set of eigenstates of the finite-time
evolution operator into the expression for the expectation value.
For time-independent rates this shows that the real part of the spectrum 
is the set of all inverse relaxation times of the process. For a stochastic 
transition matrix $H$ the real part of all eigenvalues is non-negative. A 
non-vanishing imaginary part 
signals the presence of currents characterizing an non-equilibrium system.

For time-dependent rates the probability conservation
condition $\bra{s} \, H = 0$ is also valid for all times.
However, the distribution $\ket{P^\ast}$
with eigenvalue $0$ of $H$ is time-dependent and hence not stationary.
We refer to the time-dependent distribution satisfying
$H \, \ket{P^\ast(t)} = 0$ as quasi-stationary.
We stress that (\ref{2-6}) and (\ref{2-7}) may also be satisfied for all
times and, in that case, we use the term ``time-dependent detailed balance''.

Returning to the case of time-independent rates, we remark that one can
always define an adjoint stochastic evolution $H^{\rm ad}$ by
\bel{2-7b}
H^{\rm ad} = P^\ast H^T (P^\ast)^{-1}
\ee
In the absence of detailed balance this defines a new process $H^{\rm ad} \neq H$
which has the same stationary distribution, the same waiting time distribution
for all states, and the same allowed transitions as 
$H$, but different and often
complicated non-local transition rates 
\bel{2-7d}
w^{\rm ad}_{\sigma',\sigma} = w_{\sigma,\sigma'} \frac{P^\ast(\sigma')}{P^\ast(\sigma)}
\ee
which can be computed
explicitly only if the stationary distribution is known explicitly.
The adjoint process manifestly depends on the stationary distribution
and describes a reversed time-evolution compared
with the evolution determined by $H$ since
\bel{2-7c}
\exval{A_n(t_n)\dots A_2(t_2) A_1(t_1)}^\ast = 
\exval{A_1(t-t_1) A_2(t-t_{2}) \dots A_n(t-t_n)}_{\text{ad}}^\ast
\ee
This adjoint evolution is
\emph{not} to be confused with the backward processes studied below
which are defined by the time-reversed protocol
without reference to the stationary distribution. Relation
(\ref{2-7b}) is readily extended to the time-dependent case, defining
a time-dependent adjoint evolution with time-dependent rates (\ref{2-7d}).

\subsection{Counting processes}
\label{ss:counting}

\subsubsection{Linear unidirectional coupling}

In the previous setting the evolution of the particle system was regarded
as driven by some random process that enters only implicitly through the
rates and waiting-time distribution for transitions between states.
An interesting scenario is a stochastic process that is driven by
another autonomous and explicitly given
stochastic dynamics. In other words, an underlying process 1 generates
the transitions of process 2, but process 2 has no influence on the
transitions of process 1. Then, even if process 1 is Markovian, the driven
process 2 is in general non-Markovian. 

A well-known example of such a unidirectional coupling is the passive
scalar problem in fluid dynamics where a particle (the passive scalar)
moves with the flow field of the fluid without disturbing it. Tracking
such a passive scalar provides a means for measuring properties of the
flow field.  A simple example arising naturally in lattice-gas models
is the evolution of the time-integrated particle current $J_{ij}(t)$
across a bond between sites $(i,j)$ in the lattice. In this case,
process 1 is an interacting particle system as described in the
previous subsection while process 2 is an integer-valued counting
process where the value $J_{ij}(t) \in \bbbz$ of the counter is a
stochastic variable that is increased by one unit if a particle hops
from site $i$ to site $j$ and is decreased by one unit when a particle
jumps from $j$ to $i$.

The time-integrated particle current provides an example of what we shall
call a \emph{counting process}, defined by the following two properties:\\

\noindent \emph{Property 1:} The current value $J$ of the counter does not 
influence the underlying process (unidirectional coupling,
no $J$-dependence of these rates)\\ 

\noindent \emph{Property 2:}  The value of $J$ changes only at a transition
of the underlying process with an increment that
may depend on the underlying transition, but not on the current value of $J$ 
(linearity, no $J$-dependence of the increment).\\ 

The physical scenario described by a counting process is the measurement
of some quantity $J$ by adding its increments whenever the underlying transition that
is associated with the increment occurs and under the assumption that the
measurement does not perturb the dynamics of the process on which the 
measurement is performed (at least within
the approximation of the model).
Processes of this kind arise not only for the integrated current,
but generally whenever a transition
in the particle system is associated with a change in some other
independent physical quantity such as the energy of a heat reservoir.

Such coupled processes are amenable to analysis in the
quantum Hamiltonian framework. We introduce the state space $\mathbb{J}$ of process
2 and label its states as $J$. 
Then the state space of the coupled process is $\hat{\mathbb{V}} = \mathbb{V}
\times \mathbb{J}$ and the construction becomes analogous to the
previous set-up.  A natural basis
for the coupled process is the tensor basis with basis vectors
$\ket{\sigma,J} = \ket{\sigma}\otimes \ket{J}$. The summation vector
may then be written $\bra{\hat{s}} = \bra{s}\otimes\bra{\tilde{s}}$ in terms of
the summation vectors of the two subprocesses. For independent initial
distributions of the two subprocesses the 
initial joint distribution takes the form 
$\ket{\hat{P_0}} = \ket{P_0} \otimes \ket{J_0}$.
We denote the generator of such a counting process by $\hat{H}$.

\subsubsection{Integer-valued counting processes}
\label{sss:intcount}

For integer-valued increments a counting process obviously has integer 
state space $\bbbz$ and we denote its value by $m$. If $m$
can change by only one unit in each transition of the
underlying process, e.g., if $m$ counts single particle jumps,
the transition matrix of the coupled process takes a simple form.
A single transition $\sigma \to \sigma'$ is 
generated by the matrix
\bel{2-8}
\hat{H}(\sigma',\sigma) = D(\sigma',\sigma) - 
E(\sigma',\sigma)\otimes A^\pm.
\ee
Here $A^\pm \, \ket{m} =
\ket{m\pm 1}$, and $\bra{m} \, A^\pm = \bra{m\mp 1}$,
with the sign depending on the direction of the transition 
$\sigma \leftrightarrow \sigma'$, and $D,E$ as defined in (\ref{2-4a}).\footnote{More 
generally, for a multi-step transition $m\to m'$ the off-diagonal part of the 
full evolution operator $\hat{H}$ has the form $-w_{\sigma',\sigma} \,
\ket{\sigma'}\bra{\sigma} \otimes \ket{m'}\bra{m}$ and a corresponding diagonal 
part to ensure conservation of probability.} Notice that because of the linearity
of the counting process
\bel{2-8a} 
\bra{\tilde{s}}\,A^\pm
= \bra{\tilde{s}}
\ee
which gives the simple form of the diagonal part in (\ref{2-8}).
The counting process 2 may be regarded as a non-Markovian
simple random walk whose increments have temporal correlations determined
by the underlying particle dynamics of process 1. Notice that
generically there is no joint 
stationary distribution for the
coupled process since the transition rates do not depend on
$m$ and hence there is no mechanism that limits the growth
of $m$. 

Often one is interested in information only about the counting process
and wishes to compute the generating function 
\bel{2-9}
\exval{\rme^{-\lambda m}} = \bra{\hat{s}} \, \rme^{-\lambda m} T \{ \rme^{-\int_0^t
\hat{H}(\tau) \, \rmd \tau} \} \, | \hat{P}_0 \rangle,
\ee
starting from some specific value
$m_0 \in \bbbz$ and averaged over some initial distribution $P_0$ of the
particle system, i.e., for $| \hat{P}_0 \rangle = \ket{P_0} \otimes \ket{m_0}$.
One finds
\bel{2-10}
\exval{\rme^{-\lambda m}} = \rme^{-\lambda m_0} \bra{s} \, T \{ \rme^{-\int_0^t
\tilde{H}(\tau) \, \rmd \tau} \} \, | {P}_0 \rangle
\ee
which is a scalar product on the vector space $\mathbb{V}$ spanned by the
basis vectors of the particle system only.  Here $\tilde{H}$ is a
matrix acting on $\mathbb{V}$ and obtained from $\hat{H}$ [cf.~(\ref{2-8})]
by replacing the transition matrix $A^\pm$ by the numerical factor
$\rme^{\mp \lambda}$. For a more general counting process where the
integer change $q_{\sigma',\sigma}$ in $m$ under a transition $\sigma
\to \sigma'$ depends on the transition in a more complicated fashion,
the matrix $A^\pm$ in (\ref{2-8}) has to be replaced by a matrix
$A^{q_{\sigma',\sigma}}$ with the property $A^{q_{\sigma',\sigma}}
\, \ket{m} = \ket{m+q_{\sigma',\sigma}}$. The exponential factors in
$\tilde{H}$ are then $\rme^{-\lambda q_{\sigma',\sigma}}$. Notice
that (\ref{2-10}) is a simple consequence of the fact that
$\rme^{-\lambda m}$ is a diagonal matrix that commutes with the
transition matrices $H(\sigma',\sigma)$ of the particle system and
satisfies
\bel{2-10a}
\rme^{-\lambda m}A^\pm\rme^{\lambda m} =\rme^{\mp \lambda}A^\pm.
\ee
Specifically, this gives
\be
\rme^{-\lambda m}\hat{H}(\sigma',\sigma)\rme^{\lambda m} = D(\sigma',\sigma) - 
\rme^{\mp \lambda} E(\sigma',\sigma)\otimes A^\pm
\ee
and we can then use~\eref{2-8a} to remove the operator $A^\pm$
\be
\bra{\hat{s}} [D(\sigma',\sigma) - 
\rme^{\mp \lambda} E(\sigma',\sigma)\otimes A^\pm] = \bra{\hat{s}} [D(\sigma',\sigma) - 
\rme^{\mp \lambda} E(\sigma',\sigma)]
\ee
and obtain~\eref{2-10}.
Clearly 
$\exval{\rme^{-\lambda m}}$ depends also on the initial distribution
$P_0$ of the underlying process 1. Where required
we denote this dependence of expectation values on the initial
distribution by a subscript $\exval{\rme^{-\lambda m}}_{P_0}$.
For $m_0$ we choose without loss of generality $m_0=0$.

We remark that $\tilde{H}$ is not a stochastic generator which conserves
probability and has left eigenvector $\bra{s}$ with eigenvalue 0. 
Nevertheless the evolution under $\tilde{H}$ has a straightforward
stochastic interpretation:
each stochastic trajectory generated by 
the underlying process $H$ gets weighted by
a factor $\rme^{-\lambda q_{\sigma',\sigma}}$ whenever a transition
$\sigma \to \sigma'$ occurs. Hence the expression analogous to the
conditional probability (\ref{2-5a}) gives 
the sum of the weighted trajectories from a
configuration $\sigma$ to a configuration $\sigma'$ for a time
interval of length $t$.

Since counting processes may relate to the interaction of the particle
system with its physical environment, which enters (implicitly) in the transition
rates, it is meaningful to generalize the previous setting slightly.
We note that the same transition $\sigma\to\sigma'$ may be activated by
different physical mechanisms, each of which occurs with a rate
$w_{\sigma',\sigma}^{(\alpha)}(t)$. The total rate of transition
is then $\sum_\alpha w_{\sigma',\sigma}^{(\alpha)}(t)$, but 
one may want to keep track of each process individually. 
One introduces several counters and the value of
the counter $\alpha$ is only changed under a transition of type
$\alpha$.  The corresponding Hamiltonian would then have a sum
$-\sum_\alpha w_{\sigma',\sigma}^{(\alpha)}(t)
A^{q_{\sigma',\sigma}^{(\alpha)}}_{\alpha} \, \ket{\sigma'}\bra{\sigma}$
as off-diagonal elements where the transition matrix
$A^{q_{\sigma',\sigma}^{(\alpha)}}_{\alpha}$ changes only the counter
$\alpha$.  An example would be exchange of particles on the same
lattice with different external reservoirs when one is interested in
the contribution to the total change in particle number from each reservoir
separately so that one can compute or measure joint expectation
values.

Alternatively, if transitions of different types increment the
\emph{same} counter but by \emph{different} amounts, one has a
modified Hamiltonian with off-diagonal elements $-\sum_\alpha
w_{\sigma',\sigma}^{(\alpha)}(t) A^{q_{\sigma',\sigma}^{(\alpha)}} \,
\ket{\sigma'}\bra{\sigma}$ (corresponding to setting all $\lambda$'s
in the joint generating function equal).  In other words, this is a
single counting process where the increment for a given transition
depends on the underlying mechanism.  For an illustration of how this
works in practice, see the toy example in \ref{s:app} where we weight
trajectories involving particle moves on/off a single lattice site
differently according to whether the particle is exchanged with the
left-hand reservoir or the right-hand reservoir.

\subsubsection{Real-valued counting processes}

Although the notion of ``counting'' naturally suggests a process with
integer state space, it is straightforward to extend the discussion of
the previous subsection to real-valued processes.  In slight abuse of
language, we continue to call these ``counting processes''; indeed in
the remainder of the review we shall use the term without qualification
to include this more general case.

Specifically, a real-valued counting process is defined in such a way
that a stochastic variable $x\in\bbbr$ is incremented by an amount
$r_{\sigma',\sigma}$ where $r\in\bbbr$ depends on the transition
$\sigma\to\sigma'$. 

Such processes may have continuous state space; in
this case they are constructed in a fashion similar to discrete
counting processes with the shift operator $\exp{(r\partial_x)}$
acting on basis vectors as $\exp{(r\partial_x)} \, \ket{x} = \ket{x+r}$
taking the role of $A^+$ and $\bra{x}\,x'\,\rangle = \delta(x-x')$.
Correspondingly one has $\bra{x} \exp{(r\partial_x)} =\bra{x-r}$.
The analogue of (\ref{2-10a}) reads
\bel{2-10b}
\rme^{-\lambda x} \rme^{r\partial_x} \rme^{\lambda x} 
=\rme^{\mp \lambda} \rme^{r\partial_x}
\ee
Notice that $\bra{\tilde{s}} = \int_\bbbr  dx \bra{x}$ and therefore
$\bra{\tilde{s}} \exp{(r\partial_x)} = \bra{\tilde{s}}$ in complete
analogy to the corresponding relations for integer counting processes.

In general, the amount added to $x$ when a certain transition occurs
may also depend explicitly on the time at which that transition takes
place.  Where necessary, we write $r_{\sigma',\sigma}(t)$ to make this
dependence clear.

\subsection{Functionals over trajectories}
\label{ss:functionals}

The fluctuation theorems presented in this review are concerned with
the probability distributions of trajectory-dependent functionals.
Here we introduce the particular type of functionals we will be
interested in and demonstrate their intimate connection to the
counting processes of the preceding subsection.

\begin{figure}
\begin{center}
\includegraphics*[width=0.5\textwidth]{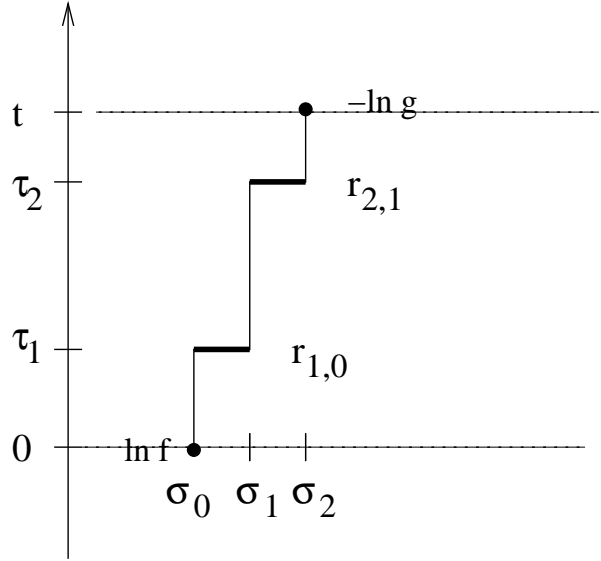}
\caption{Contribution to the functional (\ref{e:X}) for 
a stochastic trajectory $\{\sigma_0,\sigma_1,\sigma_2\}$. 
Time points upwards, the horizontal
direction is the abstract space of configurations.
The time arguments are suppressed, the
weighting factor $r_{\sigma_i, \sigma_j}(\tau)$ is abbreviated
as $r_{i,j}$. Hence $\mathcal{X}_F = \ln f_0 + r_{1,0}(\tau_1)+r_{2,1}(\tau_2) 
- \ln g_2$ and 
$\mathcal{X}_B = \ln g_0 + r_{1,0}(t-\tau_1) + r_{2,1}(t-\tau_2)
- \ln f_2$.}
\label{f:functional}
\end{center}
\end{figure}

In general, the value of a functional of a trajectory at time $t$
depends on the complete history of the Markov process until that
point, i.e., on $\{\sigma\} \equiv \{ \sigma(\tau),0 \leq \tau \leq t
\}$.  Here we consider a restricted class of functionals which have an
explicit dependence only on the ``jumps'' between configurations and
on the initial and final states of the system. 
(i.e., $\sigma(0)\equiv\sigma_0$ and $\sigma(t)\equiv\sigma_n$).  
To be more concrete, we are interested in functionals of the
form~\footnote{We will often suppress notationally the arguments of the
functional but the ``script'' letter $\mathcal{X}$ serves as a
reminder that its value depends on the history of the process.}
\begin{equation}
\fl \mathcal{X}_F(t,\{\sigma\},f,g)
= \int_0^t \sum_{\sigma,\sigma'} r_{\sigma', \sigma}(\tau) 
\Delta_{\sigma', \sigma} (\tau) \, \rmd \tau 
+ \ln[f(\sigma(0))] - \ln[g(\sigma(t))] \label{e:X}
\end{equation}
where $f$ and $g$ are arbitrary positive functions acting on the state
space. For notational clarity, we stress that the third argument of the
functional is a function of the initial value of the trajectory
$\{\sigma\}$ and the fourth argument is a function of the final value 
of the trajectory. The logarithms and signs are included for later notational
convenience. $\Delta_{\sigma', \sigma}(\tau)$ is a sum of delta
functions $\delta(\tau-\tau_k)$ located at the random times $\tau_k$
when $\sigma(\tau)$ changes from $\sigma$ to $\sigma'$.  Each delta
function is weighted by $r_{\sigma', \sigma}(\tau)$, which can be
thought of as the amount of some quantity (charge, mass, energy)
transferred in moving from configuration $\sigma$ to $\sigma'$ at time
$\tau$.  Note that by writing the sequence of delta functions
explicitly
\begin{equation} 
\Delta_{\sigma', \sigma} (\tau) = \sum_{k=1}^{n}
\delta(\tau-\tau_k) \delta_{\sigma',\sigma_k}
\delta_{\sigma,\sigma_{k-1}}
\end{equation}
we can equivalently write the first term of~\eref{e:X} as
\begin{equation} 
\sum_{k=1}^{n}
r_{\sigma_{k}, \sigma_{k-1}}(\tau_{k}).
\end{equation} 
It is now immediately clear that this term arises from a counting
process as discussed above.  The functional~\eref{e:X} is the result
of an operation where the increments $r$ of some physical quantity
associated with each transition $\sigma \to\sigma'$ are measured and
summed up and some physical quantities $\ln{f},-\ln{g}$ that depend on
the initial and final states of the system are added (see
figure~\ref{f:functional}).  These boundary values may correspond to
intrinsic observables of the configuration such as particle number,
energy, and so on, or they may represent functions of the statistical
weight assigned to the initial and final endpoints of a specific
trajectory as determined by the experimental conditions of the
measurement. Hence the functional $\mathcal{X}_F$ of the trajectory is
the \emph{value} of a counting process at time $t$ defined by the
increments $r_{\sigma', \sigma}(\tau)$ for a transition $\sigma \to
\sigma'$ at time $\tau$ (the ``hopping part'' of the functional) plus
the terminal values $\ln[f(\sigma(0))] - \ln[g(\sigma(t))]$ (the
``boundary part'').  For a given stochastic process this is a random
variable that is uniquely determined by the trajectory $\{\sigma\}$
the system has taken. Therefore the counting process expresses a
quantity that is non-local in time (the hopping part of the
functional) in terms of a quantity that is local in time (the value of
the counter at time $t$).

We shall also consider a backward functional $\mathcal{X}_B$
with weights at time $\tau$ replaced by weights at time
$t-\tau$, and the role of $f$ and $g$ interchanged.
Hence 
\begin{equation}
\fl \mathcal{X}_B(t,\{\sigma\},g,f)
= \int_0^t \sum_{\sigma,\sigma'} r_{\sigma', \sigma}(t-\tau) 
\Delta_{\sigma', \sigma} (\tau) \, \rmd \tau 
+ \ln[g(\sigma(0))] - \ln[f(\sigma(t))] \label{e:XB}
\end{equation}
As above, the third (fourth) argument is a function evaluated at the
initial (final) point of the trajectory.

We emphasize that the quantities $r_{\sigma',
\sigma}(\tau)$ can take (in principle) \emph{any} real form.  In the particular case
in which they are antisymmetric $r_{\sigma', \sigma} = - r_{\sigma,
\sigma'}$ (i.e., the weight of the reverse move is $-1$ times that of
the forward move), then the quantity measured by the counting process
has a meaningful interpretation as an integrated current
$\mathcal{J}_r$ associated with $r$. This is especially clear if $r$ is
time-independent since then we have
\begin{equation} 
\mathcal{J}_r(t,\{\sigma\})=\sum_{\sigma, \sigma'}
r_{\sigma', \sigma} J_{\sigma', \sigma} \label{e:Jr}
\end{equation} 
where $J_{\sigma', \sigma}$ is the integrated transition current (net number
of transitions) between states
$\sigma$ and $\sigma'$ along the trajectory.  For example, in a
one-dimensional nearest-neighbour exclusion process, where the only allowed transitions are
single-particle hopping events and with $r=\pm 1$ chosen to depend only on
the hopping direction across a bond in a given transition, then~\eref{e:Jr} gives
the sum of the particle currents across each bond and hence the total
integrated particle current. 

It will transpire that such currents have
a natural thermodynamic interpretation even far from equilibrium. Hence
for antisymmetric $r$
we make the nature of the two contributions to the functional
$\mathcal{X}_F$ explicit by writing it in the form
\begin{equation} 
\mathcal{X}_F(t,\{\sigma\},f,g) = \mathcal{J}_r(t,\{\sigma\})
+ \mathcal{B} (f,g)\label{e:fparts}
\end{equation} 
with boundary part $\mathcal{B}$ and hopping part $\mathcal{J}_r$
(for antisymmetric increments we refer to this as the ``current part''). 
The functional $\mathcal{X}_B$ can be split analogously.
For brevity and where no confusion can arise we often simply say current
instead of integrated current.

The distribution of a functional $\mathcal{X}$ of the form (\ref{e:X})
or (\ref{e:XB}) over an ensemble of
trajectories (defined by the initial distribution and subsequent
dynamics) is the probability that the measured value $\mathcal{X}$,
averaged over histories and initial configurations, takes some 
specific value. It is characterized by the generating function
$\langle \, \rme^{-\lambda \mathcal{X}} \, \rangle$. Since the hopping part of
$\mathcal{X}$ is also
the value of the random variable measured by the corresponding
counting process at time $t$ we can express the ensemble average 
over trajectories by an ensemble average over configurations of 
the unidirectionally-coupled joint process at time $t$.
 Including the boundary terms
and using commutation relations
within the quantum Hamiltonian formalism, one can show in analogy to
the proof of (\ref{2-10}) that this generating function can be
expressed as 
\begin{equation} 
\langle \, \rme^{-\lambda \mathcal{X}} \, \rangle_{P_0} = \langle \,
s \, | \, g^{\lambda} T \{ \rme^{-\int_0^t \tilde{H}_r(\lambda,\tau) \, \rmd
\tau} \} f^{-\lambda} \, | \, P_0 \, \rangle.
\end{equation} 
Here $f$ and $g$ are the diagonal matrices with elements $f(\sigma)$
and $g(\sigma)$ respectively and $\tilde{H}_r(\lambda,\tau)$ is
constructed from $H(\tau)$ by replacing the off-diagonal elements
$-w_{\sigma', \sigma}(\tau)$ with $-w_{\sigma',
\sigma}(\tau)\rme^{-\lambda r_{\sigma', \sigma}(\tau)}$.  Note that
for $\lambda \neq 0$, $\tilde{H}_r(\lambda,\tau)$ is no longer a
stochastic Hamiltonian (i.e., $\bra{s} \, \tilde{H}_r \neq 0$ ) but, for
particular choices of $r$, it has important symmetry properties, which
underly the fluctuation theorems of the following sections.

We conclude this subsection by remarking that the present approach is
easily extended to the calculation of joint distributions---in that
case one finds
\begin{equation} 
\langle \, \rme^{-\lambda_a \mathcal{X}_a-\lambda_b
\mathcal{X}_b} \, \rangle_{P_0} = \langle \, s \, | \, g_a^{\lambda_a} g_b^{\lambda_b} T \{
\rme^{-\int_0^t \tilde{H}(\lambda_a,\lambda_b,\tau)}
\, \rmd \tau \} f_a^{-\lambda_a} f_b^{-\lambda_b}\, | \, P_0 \, \rangle. \label{e:joint}
\end{equation} 
Here
$\tilde{H}(\lambda_a,\lambda_b,\tau)$ has off-diagonal
elements $-w_{\sigma', \sigma}(\tau)\rme^{-\lambda_a r^{(a)}_{\sigma',
\sigma}(\tau)-\lambda_b r^{(b)}_{\sigma', \sigma}(\tau)}$, as follows
from the discussion of~\ref{sss:intcount} for the case of several
distinct counters.

\section{Key concepts}
\label{s:key}

Fluctuation theorems are concerned with symmetries in the probability
distributions of certain functionals.  Loosely
speaking, the relevant functionals arise from comparing the
probability of a trajectory in phase-space to the probability of a
``time-reversed'' trajectory.  In the following we make this notion
concrete before deriving, in section~\ref{ss:basic}, a single general symmetry relation
from which specific fluctuation relations that have appeared in the literature
arise. This approach provides us with a unifying description of the various
fluctuation theorems available for stochastic dynamics.

\subsection{Time reversal and associated functionals}
\label{sss:time}

The notion of time reversal is central to the development of fluctuation 
theorems \citeaffixed{Maes03}{see e.g.,}. We emphasize here that it is crucial to distinguish
between two different senses of ``time reversal'', namely a reversal of the 
stochastic dynamics as determined by the
adjoint dynamics (\ref{2-7b}) and a reversal of
the protocol. The latter simply means that, if the rates under the forward protocol are
$w^F_{\sigma', \sigma}(\tau)$, then the rates under the backward protocol
are given by $w^B_{\sigma', \sigma}(\tau) = w^F_{\sigma',
\sigma}(t-\tau)$ [i.e., $l^B(\tau)=l^F(t-\tau)$].  In order to keep the
distinction clear we shall use the term \emph{backward protocol} or
\emph{backward process} for time reversal of the protocol while the
notion \emph{time-reversed} (or \emph{adjoint}) \emph{process} is
reserved for time reversal of the dynamics. See
\citeasnoun{Chernyak06} for a related discussion of these two concepts
within a Langevin path-integral analysis.  We will also need the
notion of a time-reversed trajectory as defined by
\bel{histrev}
\{ \sigma
\}_{\text{rev}} \equiv \{ \sigma(t-\tau), 0 \leq \tau \leq t\}.
\ee
Hence the initial (final) configuration of the time-reversed trajectory is
$\sigma_n$ ($\sigma_0$).

Fluctuation theorems arise from relationships between various 
functionals of trajectories.  Specifically, the FTs to be discussed in
the body of this review are generated by considering the
value of some counting process under reversal of the trajectories and
reversal of the protocol.  At the end of the paper (in
section~\ref{s:dis}) we will briefly discuss similar fluctuation theorems 
that arise from considering reversal of the dynamics.

We start by defining carefully in
probabilistic terms the central object of interest before discussing
its physical significance in the next subsection.
For now, we merely point out the ratio of the rate for a change in configuration
from $\sigma$ to $\sigma'$ to the corresponding rate from $\sigma'$ to
$\sigma$ is a measure of the irreversibility of the transition.
This motivates the introduction of the counting process with antisymmetric
increments
\begin{equation}
r^{(1)}_{\sigma',\sigma}(\tau)
=\ln \left[ \frac{w_{\sigma',\sigma}(\tau)}{w_{\sigma,\sigma'}(\tau)}
\right] \label{e:r1}
\end{equation}
for a transition $\sigma\to\sigma'$ at time $\tau$.  This counting
process is generically defined in terms of the transition rates so 
that when applied to the forward process we have
$w_{\sigma',\sigma}(\tau)=w^F_{\sigma',\sigma}(\tau)$ whereas for the
backward process we have
$w_{\sigma',\sigma}(\tau)=w^B_{\sigma',\sigma}(\tau)$.  In other words,
the counting process measures the \emph{same physical quantity} (ratio
of transition rates at the instant of measurement)
regardless of the direction of the protocol. In a more general treatment 
one may also consider the value of the counting process for the adjoint 
dynamics in which case 
$w_{\sigma',\sigma}(\tau)=w^{\rm ad}_{\sigma',\sigma}(\tau)$. This is
beyond the scope of this review.

We now combine this counting process with boundary terms (i.e.,
weights attached to the terminal configurations of the trajectory)
\begin{equation}
f(\sigma)=P_F(\sigma), \quad g(\sigma)=P_B(\sigma), \label{e:Rbound}
\end{equation}
to yield a functional of the type discussed in section~\ref{ss:functionals}:
\begin{equation} 
\fl \mathcal{R}_F(t,\{\sigma\},P_F,P_B)
=\ln \left[
\frac{w^F_{\sigma_n,\sigma_{n-1}}(\tau_n) \ldots
w^F_{\sigma_2,\sigma_1}(\tau_2) \, w^F_{\sigma_1,\sigma_0}(\tau_1)
P_F(\sigma_0)}{w^F_{\sigma_{n-1},\sigma_n}(\tau_n) \ldots
w^F_{\sigma_1,\sigma_2}(\tau_2 ) \, w^F_{\sigma_0,\sigma_1}(\tau_1)
P_B(\sigma_n)} \right], 
\label{e:Rdef}
\end{equation} 
together with the associated ``backward'' functional
\begin{equation} 
\fl \mathcal{R}_B(t,\{\sigma\},P_B,P_F)
=\ln \left[
\frac{w^B_{\sigma_n,\sigma_{n-1}}(\tau_n) \ldots
w^B_{\sigma_2,\sigma_1}(\tau_2) \, w^B_{\sigma_1,\sigma_0}(\tau_1)
P_B(\sigma_0)}{w^B_{\sigma_{n-1},\sigma_n}(\tau_n) \ldots
w^B_{\sigma_1,\sigma_2}(\tau_2) \, w^B_{\sigma_0,\sigma_1}(\tau_1)
P_F(\sigma_n)} \right]. \label{e:RBdef}
\end{equation}  
[See Figure~\ref{f:functional} for an illustration of the evaluation
of forward and backward functionals for a simple trajectory.]  Notice
that, if $\mathcal{R}_F$ and $\mathcal{R}_B$ are to be well-defined
for all possible trajectories, we require $P_B(\sigma)$ and
$P_F(\sigma)$ to be non-zero for all $\sigma$.  This is related to the
condition of ``ergodic consistency'' \cite{Evans02b}.  Below we shall
discuss how a type of fluctuation theorem can still be obtained even
if this requirement is not satisfied.

The functionals defined above are a generalization of the
``action functional'' of \citeasnoun{Lebowitz99}. Their relationship to
time reversal becomes clearer by noting that $\mathcal{R}_F$ can be written in the
following form
\begin{equation} 
\mathcal{R}_F
=\ln \left[
\frac{w^F_{\sigma_n,\sigma_{n-1}}(\tau_n) \ldots
w^F_{\sigma_2,\sigma_1}(\tau_2) \, w^F_{\sigma_1,\sigma_0}(\tau_1)
P_F(\sigma_0)}{w^B_{\sigma_0,\sigma_1}(t-\tau_1) \,
w^B_{\sigma_1,\sigma_2}(t-\tau_2) \ldots
w^B_{\sigma_{n-1},\sigma_n}(t-\tau_n) P_B(\sigma_n)} \right]. 
\label{e:Rdefb}
\end{equation} 
If we now take $P_F$ to be the initial probability distribution of the forward
process and $P_B$ to be the initial distribution of the backward
process we obtain a probabilistic meaning for the functional---in words, it is
just the logarithm of the ratio of the probability of seeing a
trajectory $\{ \sigma \}$ in the forward process (starting from an
initial distribution $P_F$) to the probability of seeing
the {\it time-reversed} trajectory $\{ \sigma \}_{\text{rev}}$ in the
\emph{backward} process (starting from $P_B$).  Note that the exponential factors containing
inverse waiting times all cancel in the probability ratio.  A
thermodynamic interpretation of the current part of the functional
(and hence of the counting process) as dissipated heat will be given
below. Here we merely remark that in processes with time-independent
rates obeying the detailed balance condition (\ref{2-6}), the current
part of $\mathcal{R}_F$ reduces to boundary terms. If then the initial
and final configurations are weighted by the equilibrium probabilities, i.e., $P_F(\sigma_0) =
P^\ast(\sigma_0)$, $P_B(\sigma_n) = P^\ast(\sigma_n)$, all boundary
terms cancel, i.e., the probability of the two trajectories is the
same and $\mathcal{R}_F=\mathcal{R}_B=0$. Hence, as indicated above,
the functionals
$\mathcal{R}_F,\mathcal{R}_B$ with judiciously chosen boundary parts
are a measure for the irreversibility of the process.

In our study of fluctuation theorems we will be interested in the
value of the functional $\mathcal{R}_F$ in a process governed by the
forward protocol compared to the value of the functional
$\mathcal{R}_B$ in the equivalent backward process.  Later we will
see how suitable choice of the boundary terms leads to the
identification of these functionals with the same physical quantity for
forward and backward processes.  In anticipation of this and
for notational simplicity we suppress the subscript inside expectation values.
Moreover, since $P_F$ ($P_B$) is always taken to be the initial distribution
of the forward (backward) process, we implicitly assume expectation values
to be evaluated under the measure generated by the forward (backward) 
evolution with initial measure $P_F$ ($P_B$).
Hence we shall write $\exval{A(\mathcal{R})}^F$ as a shorthand
for the expectation $\exval{A(\mathcal{R}_F)}_{P_F}^F$ of some
function $A$ of the forward functional under forward evolution
with initial distribution $P_F$, i.e., the expectation of 
the random number $\mathcal{R}_F$ is taken with respect to the statistical 
weights of trajectories under forward evolution, with initial 
configurations weighted by $P_F$. Analogously, we write
$\exval{A(\mathcal{R})}^B$ for the expectation 
$\exval{A(\mathcal{R}_B)}_{P_B}^B$ of the random number
$\mathcal{R}_B$.

In section~\ref{ss:basic} and following we will show how symmetries of
the generating functions for $\mathcal{R}_F$ and $\mathcal{R}_B$ underly various
forms of finite-time fluctuation relations.  First, however, we
digress to introduce the physical concept of entropy and discuss its
relation to the abstract quantities defined above.

\subsection{Entropy and heat}
\label{sss:entheat}

The entropy of a system described by stochastic dynamics 
is the usual Gibbs entropy
\bel{entropy}
S(t) = - \sum_{\sigma} P(\sigma;t) \ln P(\sigma;t) = - \exval{\ln P(\sigma;t)}.
\ee 
The role of time reversal in the construction of the functional
$\mathcal{R}_F$ hints at a connection to the concept of entropy. In
order to explore this further, we must first
discuss how to define entropy changes at the level of individual 
trajectories in phase space.  There seems to be some ambiguity about
the identification of such entropy terms. The treatment in this
subsection is largely based on that of \citeasnoun{Seifert05} but there may
be other self-consistent schemes; see also the discussion
in \citeasnoun{Lebowitz99}.

As a starting point we take the definition of \citeasnoun{Schnakenberg76} for the
ensemble-averaged rate of total entropy production in a
non-equilibrium process.\footnote{In fact,~\eref{e:ent} is a slight
generalization of Schnakenberg's original expressions for processes with 
constant transition rates.} 
In our notation this reads
\begin{equation}
\dot{S}_\text{tot}(\tau) \equiv 
 - \sum_{\sigma,\sigma'}  w_{\sigma',\sigma}(\tau) P(\sigma;\tau) \ln \left[
  \frac{w_{\sigma',\sigma}(\tau) P(\sigma;\tau)}{ 
    w_{\sigma,\sigma'}(\tau) P(\sigma';\tau)} \right]. \label{e:ent}
\end{equation}
Notice that $\dot{S}_\text{tot} \geq 0$ as required.
Now, we argue that this total entropy production (generated by the
interaction of the system with particle/heat reservoirs which leads to the
stochastic jumps) can be divided into terms representing the entropy 
production of the stochastic ``system'' itself and an entropy flow that 
leads to a change of the entropy of the environment (the reservoirs). 
This is the analogue of the distinction in Langevin dynamics between the 
entropy of the ``particle'' and that of the ``medium''. The entropy production
of the system is the time-derivative of the usual Gibbs entropy 
(\ref{entropy}), viz.,
\begin{equation}
  \dot{S}_\text{sys}(\tau)=-\sum_{\sigma,\sigma'} 
  w_{\sigma',\sigma}(\tau) P(\sigma;\tau) \ln \left[
  \frac{P(\sigma;\tau)}{P(\sigma';\tau)} \right] \label{e:sysent}
\end{equation}
and correspondingly the entropy production of the environment is given by
\begin{equation}
 \dot{S}_\text{env}(\tau)= \sum_{\sigma,\sigma'} 
 w_{\sigma',\sigma}(\tau) P(\sigma;\tau) \ln \left[
   \frac{w_{\sigma,\sigma'}(\tau)}{w_{\sigma',\sigma}(\tau)}\right]. \label{e:envent}
\end{equation}
It is straightforwardly verified that $\dot{S}_\text{tot} = 
\dot{S}_\text{sys} + \dot{S}_\text{env}$. 

These statements for ensemble averages can be given a microscopic
meaning in terms of the entropy change along a single 
trajectory \cite{Seifert05}. To this end, one goes back from the
ensemble-averaged Gibbs entropy to microscopic configurations
and assigns an ``entropy'' 
\begin{eqnarray}
\mathcal{S}_\text{sys}(\tau) &= - \ln{P(\sigma_k;\tau)} 
\quad &\text{for } \tau_k \leq \tau \leq \tau_{k+1}\\
&= - \ln{P(\sigma(\tau);\tau)} 
\quad &\text{for } 0 \leq \tau \leq t, \label{histentropy}
\end{eqnarray}
for the configurations of a trajectory 
$\{\sigma\}$.\footnote{Note that this single-trajectory quantity still 
makes implicit reference to an 
ensemble of trajectories through the probability $P(\sigma;\tau)$.} 
Here $P(\sigma(\tau);\tau)$ is the solution of the master equation
(with given initial condition) for
the configuration $\sigma(\tau)$. Hence this microscopic entropy 
undergoes a jump discontinuity whenever a transition occurs.
In analogy to the macroscopic quantities considered above
the time derivative 
\begin{equation}\label{entropyprodsys}
\dot{\mathcal{S}}_\text{sys}(\tau) = - \frac{\partial_\tau
  P(\sigma(\tau);\tau)}{P(\sigma(\tau);\tau)} - \sum_{k=1}^n \delta(\tau-\tau_k)  \ln \left[
  \frac{P(\sigma_{k};\tau_k)}{P(\sigma_{k-1};\tau_k)} \right] 
\end{equation}
can be split into two parts
\begin{eqnarray}
 \dot{\mathcal{S}}_\text{tot}(\tau) &\equiv - \frac{\partial_\tau
  P(\sigma(\tau);\tau)}{P(\sigma(\tau);\tau)} - \sum_{k=1}^n \delta(\tau-\tau_k)  \ln \left[
  \frac{w_{\sigma_{k-1},\sigma_{k}}(\tau_k)P(\sigma_{k};\tau_k)}
{w_{\sigma_{k},\sigma_{k-1}}(\tau_k)P(\sigma_{k-1};\tau_k)} \right] 
\\
 \dot{\mathcal{S}}_\text{env}(\tau) &\equiv  \sum_{k=1}^n \delta(\tau-\tau_k)  \ln \left[
  \frac{ w_{\sigma_{k},\sigma_{k-1}}(\tau_k)}
       { w_{\sigma_{k-1},\sigma_{k}}(\tau_k)} \right] 
\end{eqnarray}
with $\dot{\mathcal{S}}_{sys}=\dot{\mathcal{S}}_{tot}-\dot{\mathcal{S}}_{env}$.
Taking ensemble averages leads back to the analogous macroscopic expressions
(\ref{e:ent})--(\ref{e:envent}). Note that alternatively one can write
$S_\text{env}(\tau)= \sum_{\sigma} P(\sigma;\tau) I(\sigma,\tau)$ where 
\bel{entropyflux}
I(\sigma,\tau) = \sum_{\sigma'} 
w_{\sigma',\sigma}(\tau)  \ln \left[
  \frac{w_{\sigma',\sigma}(\tau)}{w_{\sigma,\sigma'}(\tau)}\right]
\ee
is the entropy flux introduced for constant rates by \citeasnoun{Lebowitz99}. 

With these definitions we see that the change in environment entropy along a
single trajectory 
\be
\Delta \mathcal{S}_\text{env} =
\int_0^t\,\dot{\mathcal{S}}_\text{env}(\tau) \,d\tau\
\ee
is the current part (measured by the counting process defined
by~\eref{e:r1}) which contributes to $\mathcal{R}_F$ and $\mathcal{R}_B$.  
Since for thermal systems at temperature $T$, the
dissipated heat $\mathcal{Q}= T\Delta \mathcal{S}_\text{env}$, we can
also think of this current part as a generalized non-equilibrium heat term. In other
words we write,
\begin{equation}
\mathcal{R}_F = \frac{\mathcal{Q}}{T} + \ln \left[ \frac{
P_F(\sigma_0)}{ P_B(\sigma_n)} \right]
\end{equation}
or equivalently
\begin{equation}
\mathcal{R}_B = \frac{\mathcal{Q}}{T} + \ln \left[ \frac{
P_B(\sigma_0)}{ P_F(\sigma_n)} \right]
\end{equation}
For athermal systems the quantities $\mathcal{Q}$ and $T$ cannot be
measured separately but their ratio (i.e., the entropy change of the
environment) is still a meaningful quantity.  This gives a microscopic
physical interpretation to the current term in $\mathcal{R}_F$ and
$\mathcal{R}_B$.  We defer discussion of the physical identification
of the boundary terms to section~\ref{s:finite}; at present $P_F$ and
$P_B$ should be considered arbitrary and independent.

\subsection{Fundamental fluctuation relation}
\label{ss:basic}

After the effort of setting-up the appropriate machinery (in
particular, the master equation approach with the 
counting process for keeping track
of the entropy flux along a single trajectory) the central
proofs of fluctuation relations are remarkably simple.

We start by considering the distribution of the functional
$\mathcal{R}_F$~\eref{e:Rdef} in the forward process starting from a distribution $P_F$.
If follows from the discussion in section~\ref{ss:functionals} that
the generating function is given by
\begin{eqnarray}
\langle \, \rme^{-\lambda \mathcal{R}} \, \rangle^F &= \langle \, P_B^\lambda(t) 
\rme^{-\lambda \mathcal{J}_\mathcal{Q}(t)} P_F^{-\lambda}(0) \, \rangle^F \\
&= \langle \, s \, | \, P_B^\lambda 
T\{ \rme^{-\int_0^t \tilde{H}_\mathcal{Q}(\lambda,\tau) \, \rmd \tau} \}
P_F^{-\lambda} 
\, | \, P_F \rangle \label{e:gen1} 
\end{eqnarray}
where $\tilde{H}_\mathcal{Q}(\lambda,\tau)$ is the modified
Hamiltonian for the counting process~\eref{e:r1} with \emph{forward}
rates, i.e., it has off-diagonal elements $-w^F_{\sigma',
  \sigma}(\tau)\rme^{-\lambda \ln [w^F_{\sigma',
\sigma}(\tau) / w^F_{\sigma,\sigma'}(\tau)]}$.  We assume an initial
value $ \mathcal{J}_\mathcal{Q}(0)=0$ for the current part and use the subscript
$\mathcal{Q}$ to remind the reader of the interpretation as heat (see
section~\ref{sss:entheat}).   Notice that introducing the counting
process allows us to express the expectation of the functional
$\mathcal{R}_F$ which is non-local in time by a joint expectation of
quantities that are local in time. This is the crucial trick that
makes the derivation of fluctuation relations almost trivial.

To measure $\mathcal{R}_B$ in the backward process we use instead
$\tilde{H}_\mathcal{Q}(\lambda,t-\tau)$.  For initial distribution
$P_B$ the generating function is given by
\begin{eqnarray}
\langle \, \rme^{-\lambda \mathcal{R}} \, \rangle^B  &= \langle \, P_F^\lambda(t) 
\rme^{-\lambda \mathcal{J}_\mathcal{Q}(t)} P_B^{-\lambda}(0) \, \rangle^B \\
&= \langle \, s \, | \, P_F^\lambda 
T\{ \rme^{-\int_0^t \tilde{H}_\mathcal{Q}(\lambda,t-\tau) \, \rmd \tau} \}
P_B^{-\lambda} 
\, | \, P_B \, \rangle.
\end{eqnarray}

The next step is to realize that the connection 
to time reversal is manifest in a particular symmetry of
$\tilde{H}_\mathcal{Q}(\lambda,\tau)$, viz.
\begin{equation}
{\tilde{H}}_\mathcal{Q}^T(\lambda,\tau)=\tilde{H}_\mathcal{Q}(1-\lambda,\tau).
\label{e:Hsym2}
\end{equation} 
This is straightforwardly proved by noting that, for the off-diagonal elements,
\begin{eqnarray}
\left[{\tilde{H}}_\mathcal{Q}^T(\lambda,\tau)\right]_{\sigma',\sigma}&=
\left[{\tilde{H}}_\mathcal{Q}(\lambda,\tau)\right]_{\sigma,\sigma'}
\\ &= -w^F_{\sigma,\sigma'}(\tau) \exp \left[-\lambda \ln
\frac{w^F_{\sigma, \sigma'}(\tau)}{w^F_{\sigma', \sigma}(\tau)} \right] \\
&= -w^F_{\sigma',\sigma}(\tau) \exp \left[(1-\lambda) \ln
\frac{w^F_{\sigma, \sigma'}(\tau)}{w^F_{\sigma', \sigma}(\tau)} \right] \\
&= -w^F_{\sigma',\sigma}(\tau) \exp \left[-(1-\lambda) \ln
\frac{w^F_{\sigma', \sigma}(\tau)}{w^F_{\sigma, \sigma'}(\tau)} \right]\\
&=\left[{\tilde{H}}_\mathcal{Q}(1-\lambda,\tau)\right]_{\sigma',\sigma}
\end{eqnarray} 
whereas the diagonal elements have no $\lambda$ dependence.

Finally we use the symmetry relation~\eref{e:Hsym2} to transform the
generating function for $\mathcal{R}_F$:
\begin{eqnarray} 
\langle \, \rme^{-\lambda \mathcal{R}} \, \rangle^F 
&= \langle \, s \,| \, P_B^\lambda 
T\{ \rme^{-\int_0^t \tilde{H}_\mathcal{Q}(\lambda,\tau) \, \rmd \tau} \}
P_F^{-\lambda}\, 
|\, P_F \, \rangle  \\ 
&= \langle \, P_F \, |\, P_F^{-\lambda} 
T\{ \rme^{-\int_0^t {\tilde{H}}_\mathcal{Q}^T(\lambda,t-\tau) \, \rmd \tau} \}  
P_B^{\lambda}\, |\, s\, \rangle \\ 
&= \langle \,s \,|\, P_F^{1-\lambda} 
T\{ \rme^{-\int_0^t \tilde{H}_\mathcal{Q}(1-\lambda,t-\tau) \, \rmd 
\tau} \} 
P_B^{-(1-\lambda)}\, |\, P_B \,\rangle \\ 
& = \langle\, \rme^{-(1-\lambda)\mathcal{R}}\, \rangle^B
\end{eqnarray} 
To restate, we have obtained the following relationship relating
averages of $\mathcal{R}_F$ and $\mathcal{R}_B$ over forward and backward processes:
\begin{equation}
\langle \,\rme^{-\lambda \mathcal{R}}\, \rangle^F = \langle
\,\rme^{-(1-\lambda)\mathcal{R}} \,\rangle^B. \label{e:Rsym}
\end{equation}

We now remind the reader that we required the initial distributions
$P_F$ and $P_B$ assumed implicitly in \eref{e:Rsym} to be
non-vanishing for all $\sigma$. In order to obtain meaningful
fluctuation relations when this condition is not obeyed, we observe
that one can generalize (\ref{e:Rsym}) by introducing observables $X$,
$Y$ which are arbitrary functions of the configurations
$\sigma$. Going through the same steps as above we obtain
\begin{equation}
\langle \,Y(t)\rme^{-\lambda \mathcal{R}} X(0)\, \rangle^F = \langle\, X(t)
\rme^{-(1-\lambda)\mathcal{R}} Y(0)\,\rangle^B \label{e:fundsym}
\end{equation}
Choosing $X$ and $Y$ as indicator functions on some subset of $\mathbb{V}$
one obtains the analog of the relation (\ref{e:Rsym}) for
functions $P_F,P_B$ that vanish on the complement of this subset.
In particular, choosing $X=\ket{\sigma_\text{ini}}\bra{\sigma_\text{ini}}$ 
and $Y=\ket{\sigma_\text{fin}}\bra{\sigma_\text{fin}}$ one obtains
a symmetry relation for trajectories between fixed configurations
$\sigma_\text{ini}$, $\sigma_\text{fin}$.  This gives the detailed
fluctuation theorems introduced by \citeasnoun{Jarzynski00}.

The ``fundamental symmetry relation'' (\ref{e:fundsym}) is the central
result of our discussion which, to our knowledge, has not previously
appeared in this general form.  In the next section we
demonstrate how one can recover from this relation many
specific forms of fluctuation theorem found in the literature. 

\section{Finite-time fluctuation theorems}
\label{s:finite}

\subsection{Integral fluctuation relations}
\label{ss:int}

Setting $\lambda=1$ in~\eref{e:Rsym} gives an 
``integral fluctuation relation'' \cite{Maes03b}
\begin{equation}
\langle \,\rme^{- \mathcal{R}} \,\rangle^F = 1.  \label{e:Rint}
\end{equation}
Note that setting $\lambda=0$ yields the equivalent relation for the
backward protocol starting from distribution $P_B$.  In the remainder
of this subsection we drop the sub- and superscripts and assume for
definiteness the forward evolution.

\Eref{e:Rint} holds for $\mathcal{R}_F$ as defined in~\eref{e:Rdef} 
with \emph{any} normalized choice of $P_F$ and $P_B$ 
(recall that $P_F$ is here identified with the
initial distribution of the process over which the average is
taken).  
The specific choice of $P_F$ and $P_B$ then determines the physical interpretation
of $\mathcal{R}_F$ as we shall discuss in the following subsections.

\subsubsection{Jarzynski equality}
\label{sss:Jarzynski}

Let us take a process in which the rates (determined by the protocol
$l(\tau)$) obey the
time-dependent detailed balance condition (cf.~\eref{2-6}) with
quasi-stationary distribution $P^\text{eq}_{l(\tau)}$.  Now we imagine
preparing an experiment in which we start with initial
distribution, 
\begin{equation}
P_F(\sigma)=P^\text{eq}_{l(0)}(\sigma) \label{e:Peq1},
\end{equation}
then let the system evolve (note that in general it will not be in the
quasi-stationary distribution for $\tau > 0$) and measure a quantity
defined by $\mathcal{R}_F$~\eref{e:Rdef} with
\begin{equation}
P_B(\sigma)=P^\text{eq}_{l(t)}(\sigma) \label{e:Peq2}.
\end{equation}
In this case $\mathcal{R}_F$ is proportional to the dissipated work, 
which can be seen as follows.

Using the Boltzmann form
$P^\text{eq}_{l}(\sigma)=\rme^{-U_l(\sigma)}/Z_l$, we can write the
boundary part of the functional
$\mathcal{R}_F$ as
\begin{eqnarray}
\ln P_F(\sigma_0) - \ln P_B(\sigma_n) &= \ln
P^\text{eq}_{l(0)}(\sigma_0) - \ln P^\text{eq}_{l(t)}(\sigma_n) \\
& = \frac{\Delta U}{T} + \Delta(\ln Z).
\end{eqnarray}
Combining this with the interpretation of the current part
$\mathcal{J}_{\mathcal{Q}}$ as proportional to heat (see
section~\ref{sss:entheat}) and using the expression for free energy,
$F= - T \ln Z$, gives
\begin{equation}
\mathcal{R}_F = \frac{\mathcal{Q}}{T} + \frac{\Delta U}{T} - \frac{\Delta F}{T}.
\end{equation}
Finally, using the first law of thermodynamics we can express
$\mathcal{R}_F$ in terms of the work $\mathcal{W}=(\mathcal{Q}+\Delta U)/T$ or the dissipated
work $\mathcal{W}_d$.
\begin{eqnarray}
\mathcal{R}_F &= \frac{\mathcal{W}-\Delta F}{T} \\
&= \frac{\mathcal{W}_d}{T}.
\end{eqnarray} 
[An analogous argument shows that for the backward process with
$P_B$ as initial distribution, $\mathcal{R}_B = \mathcal{W}_d$.]

Hence, with the choices~\eref{e:Peq1} and~\eref{e:Peq2} for $P_F$ and $P_B$, relation~\eref{e:Rint} becomes the
Jarzynski equality \cite{Jarzynski97}
\begin{equation}
\langle \,\rme^{-\mathcal{W}_d/T} \,\rangle = 1. \label{e:Jar}
\end{equation}
Since $\Delta F$ (the
equilibrium free
energy difference between initial and final states) is a
trajectory-independent quantity, it can be taken 
outside the stochastic average to
yield the more usual form
\begin{equation}
\langle\, \rme^{-\mathcal{W}/T}\, \rangle = \rme^{-\Delta F / T}.
\end{equation}
This is an important statement since it implies we can measure
equilibrium free energies from an average of the non-equilibrium work
performed; see section~\ref{s:exp} for a discussion of related
experiments.  The Jarzynski equality can also be related to some
earlier work theorems \cite{Bochkov77,Bochkov79,Bochkov81}; a recent
discussion of the connections can be found in \citeasnoun{Jarzynski06b}.

Although we start from the equilibrium distribution corresponding to
$l(0)$, we emphasize that it is nowhere assumed that the system has
relaxed back to equilibrium at the final values of the rates.

\subsubsection{Integral fluctuation theorem for entropy}
\label{sss:intent}

Let us now consider a different experimental scenario.  We start the
system from some arbitrary initial distribution $P_F$ and measure the
quantity $\mathcal{R}_F$ defined by~\eref{e:Rdef} with $P_B$ chosen to
correspond to the final probability distribution of the process, i.e.,
\begin{equation}
|\,P_B\,\rangle = T \{ \rme^{-{\int_0^t H(\tau)}} \}\,|\,P_F\,\rangle \label{e:Pcon}.
\end{equation}
(for a given experiment $P_B$ is obtained empirically by repeating the
experiment many times and recording the final configuration).

Now, if $P_B$ is determined by the relation~\eref{e:Pcon}, then the
boundary term of $\mathcal{R}_F$ can be written as
\begin{equation}
\ln P_F(\sigma_0) - \ln P_B(\sigma_n) = \ln P(\sigma(0),0) - \ln P(\sigma(t),t)
\end{equation}
where $P(\sigma,t)$ is the solution of the master equation.
Comparison with the definition~\eref{histentropy} shows that these
boundary terms can be interpreted as the change in ``system'' entropy
$\Delta \mathcal{S}_\text{sys}$.  Hence in this case we have,
\begin{equation}
\mathcal{R}_F= \Delta \mathcal{S}_\text{env} + \Delta
\mathcal{S}_\text{sys} = \Delta\mathcal{S}_\text{tot},
\end{equation}
and~\eref{e:Rint} becomes an integral relation for the \emph{total}
entropy change \cite{Seifert05} 
\begin{equation}
\langle \,\rme^{-\Delta \mathcal{S}_\text{tot}} \,\rangle = 1 \label{e:ints}.
\end{equation}
Since $\rme^{\langle \, x \,\rangle} \leq \langle \,\rme^x \,\rangle$
(Jensen's inequality),~\eref{e:ints} implies $\langle \, \Delta
\mathcal{S}_\text{tot} \, \rangle \geq 0$ in agreement
with~\eref{e:ent}.  In other words, the fluctuation theorem is
entirely consistent with the Second Law of Thermodynamics which,
properly interpreted, is a statement about averages, not individual
trajectories.

The relation~\eref{e:ints} is valid for any choice of dynamics (i.e.,
with or without time-dependent detailed balance).  Notice however,
that if time-dependent detailed balance \emph{does} hold, and the system
starts in equilibrium at $l(0)$ \emph{and} is permitted to relax back
to equilibrium at $l(t)$, then~\eref{e:Jar} and~\eref{e:ints} become
identical.  For dynamics \emph{without} time-dependent detailed balance, a
special case, considered by several authors is the transition between
steady states.  In other words the system starts in the
quasi-stationary distribution at $\tau=0$ and is perturbed in such a
way as to reach stationarity at $\tau=t$ (but not necessarily at
intermediate times).  This corresponds to the choice
\begin{equation}
P_F(\sigma)=P^*_{l(0)}(\sigma) \qquad \text{and} 
\qquad P_B(\sigma)=P^*_{l(t)}(\sigma). \label{e:qs}
\end{equation}
We will return to this particular situation in section~\ref{s:dis}.

\subsection{Generating-function symmetries}
\label{ss:transient}

In the previous subsection we considered the special case $\lambda=1$.
For general $\lambda$ the full generating-function symmetry leads to various
``stronger'' fluctuation theorems which we now discuss.
We first write the
generating-function relation~\eref{e:Rsym} as
\begin{equation}
\sum_R \text{Prob}^F(\mathcal{R}_F=R) \rme^{-\lambda R} 
= \sum_R \text{Prob}^B(\mathcal{R}_B=R) \rme^{-(1-\lambda) R}. \label{e:det1}
\end{equation}
where $\text{Prob}^F(\mathcal{R}_F=R)$ denotes the probability
$\mathcal{R}_F=R$ in the forward process (with initial distribution
$P_F$) and analogously for the backward process. 
For notational simplicity, we suppress the
explicit time-dependence of these probability distributions.
This is trivially equivalent to
\begin{equation}
\sum_R \text{Prob}^F(\mathcal{R}_F=R) \rme^{-\lambda R} 
= \sum_R \text{Prob}^B(\mathcal{R}_B=-R) \rme^{(1-\lambda) R} \label{e:det2}
\end{equation}
and hence to satisfy it for all $\lambda$ requires
\begin{equation}
\frac{\text{Prob}^B(\mathcal{R}_B=-R)}{\text{Prob}^F(\mathcal{R}_F=R)}=\rme^{-R}. \label{e:transa}
\end{equation}

For the physical interpretation of this relation, we recall that $P_F$
and $P_B$ are arbitrary positive distributions, defined by physical
quantities (e.g., macroscopic observables such as internal energy,
magnetization, ...) characterizing the microstates $\sigma$.  It
follows that, in general, for $P_F$ and $P_B$ independent, the
functional $\mathcal{R}_F$ is the value of a \emph{different} quantity
in the forward process to that given by $\mathcal{R}_B$ in the
backward process.  Indeed, if $\mathcal{R}_F$ and $\mathcal{R}_B$ are
to represent the \emph{same} physical quantity we must choose the
boundary distributions to depend on the protocol in such a way that
$P_F$ becomes $P_B$ under reversal of the protocol (and vice versa).

To illustrate this, we consider the case where the distribution
$P_F$ ($P_B$) is determined by the transition rates of the forward
process at time 0 ($t$). We can write this as
\begin{equation}
P_F(\sigma)=u_{l^F(0)}(\sigma), \quad
P_B(\sigma)=u_{l^F(t)}(\sigma),  \label{e:up0}
\end{equation}
where $u_{l^F(\tau)}$ is a normalized distribution function
parameterized by the value of the protocol at time $\tau$.  In other
words, $P_F$ depends in the \emph{same} way on the rates at $\tau=0$ as $P_B$
depends on the rates at $\tau=t$. It is clear that if the distributions $P_F$ and $P_B$ obey
\eref{e:up0}, then we can associate them with an
explicitly time-dependent physical quantity $\Xi(\sigma,\tau)$ so that
for a trajectory $\{ \sigma \}$ in the forward process we have
\begin{equation}
P_F(\sigma_0)=\Xi(\sigma_0,0), \quad
P_B(\sigma_n)=\Xi(\sigma_n,t).
\end{equation}
This means that the boundary terms of $\mathcal{R}_F$ simply represent
the physical quantity $-\Delta \Xi$ for the forward process (starting
from $P_F$).  Since
$l^B(\tau)=l^F(t-\tau)$, an
analogous argument shows that the boundary part of $\mathcal{R}_B$
gives $-\Delta \Xi$ for backward evolution (starting from $P_B$).
Hence, given $r$ and $\Xi$ we can
assign to each trajectory a unique random number $\mathcal{R}$ which
takes value $\mathcal{R}_F$ ($\mathcal{R}_B$) under forward (backward)
evolution and represents the same physical quantity in each case.

We now give two concrete examples of situations
satisfying~\eref{e:up0}.  Firstly, suppose we take the choice
$u_{l^F(\tau)} = P^\text{eq}_{l^F(\tau)}$ corresponding to
\begin{equation} 
P_F(\sigma)=P^\text{eq}_{l^F(0)}(\sigma), \quad P_B(\sigma)=P^\text{eq}_{l(t)}(\sigma). 
\end{equation}
As argued in section~\ref{sss:Jarzynski}, the boundary part of the
functional $\mathcal{R}_F$ or $\mathcal{R}_B$ then gives the value of
the physical quantity $(\Delta U -\Delta F) / T$.  Secondly, consider
time-independent rates and the choice $P_F=P_B=P^*$.  In this case, the
final probability distribution of the forward process is the initial
distribution of the backward process (and vice versa) and we can
identify the boundary terms with the change in system entropy $\Delta
\mathcal{S}_\text{sys}$ (cf.\ section~\ref{sss:intent}).

The two examples mentionned above will feature prominently in the
remainder of this subsection.  More generally, we argue that in any
situation in which $P_F$ and $P_B$ are related by reversal of
protocol, $\mathcal{R}_F$ and $\mathcal{R}_B$ measure the same
physical quantity in forward and reverse processes (with initial
distributions $P_F$ and $P_B$ respectively).  We can then denote this
quantity by $\mathcal{R}$ without subscript and write~\eref{e:transa}
in the simplified form
\begin{equation}
\frac{p_B(-\mathcal{R})}{p_F(\mathcal{R})}=\rme^{-\mathcal{R}}. \label{e:trans}
\end{equation}
Here $p_F(\mathcal{R})$ denotes the probability distribution for the physical
quantity $\mathcal{R}$ in the forward process and $p_B(\mathcal{R})$ is the corresponding distribution for the
backward process. 
\Eref{e:trans} is probably the most oft-quoted form of what is known
as the transient fluctuation
theorem.  In fact it is simply a consequence of the construction of
$\mathcal{R}_F$ as the ratio of trajectory probabilities in forward and
backward processes---see~\eref{e:Rdefb}.  Note that the generating
function symmetry also
implies a relationship between the averages of any function
$A(\mathcal{R})$, viz,
\begin{equation}
\langle \, A(\mathcal{R}) \rme^{-\mathcal{R}} \,\rangle^F 
= \langle \, A(-\mathcal{R})\, \rangle^B. \label{e:usym}
\end{equation}

The subtle differences between the several transient fluctuation
theorems found in the literature (as well as their physical
interpretations) depend both on the choice of $P_F$ and $P_B$ and the
time-dependence of the protocol.  Below we discuss some particular
cases where the condition~\eref{e:up0} is obeyed and we obtain FTs of
the form~\eref{e:trans}.

\subsubsection{Crooks'  fluctuation theorem}
\label{sss:CrooksFT}

Recall from the above discussion that, for rates obeying time-dependent detailed balance,
the choice 
\begin{equation}
P_F=P^\text{eq}_{l^F(0)} \qquad \text{and} \qquad P_B=P^\text{eq}_{l^F(t)}.
\end{equation}
allows the identification of $\mathcal{R}$ as proportional to the
dissipated work $\mathcal{W}_d=\mathcal{Q}+\Delta U - \Delta F$.
\Eref{e:trans} then becomes the fluctuation theorem 
\begin{equation}
\frac{p_B(-\mathcal{W}_d)}{p_F(\mathcal{W}_d)}=\rme^{-\mathcal{W}_d/T}, \label{e:CrooksFT}
\end{equation}
which is due to \citeasnoun{Crooks99}.

\subsubsection{Evans-Searles fluctuation theorem}
\label{sss:ESFT}

For a \emph{time-independent} protocol the forward and reverse
processes are clearly identical.  If we also take $P_B=P_F$ then we
can drop the subscripts on the probability distributions
$p_F(\mathcal{R})$ and $p_B(\mathcal{R})$.  Setting $P_B=P_F=P_0$ in
the definition of $\mathcal{R}$ yields the ``dissipation function''
$\Omega$ of \citeasnoun{Evans02b}, i.e.,
$\Omega=\mathcal{R}_F(t,\{\sigma\},P_0,P_0)=
\mathcal{R}_B(t,\{\sigma\},P_0,P_0)$. 
The corresponding
fluctuation theorem \cite{Evans94,Searles99}
\begin{equation}
\frac{p(-{\Omega})}{p({\Omega})}=\rme^{-{\Omega}}. \label{e:ESFT} 
\end{equation}
holds for arbitrary initial distribution $P_0$.
It is clear that a similar transient fluctuation theorem is valid for \emph{any
time-symmetric protocol} i.e., $l(\tau)=l(t-\tau)$.

\subsubsection{Fluctuation theorem for entropy}
\label{sss:entropy}

A special case of the Evans-Searles FT corresponds to taking
$P_F=P_B=P^*$.  Experimentally, this simply means allowing a system
(with time-independent rates) to relax to stationarity before starting the
measurement. In this case
the condition~\eref{e:Pcon} is clearly obeyed and we can identify
$\mathcal{R}$ with the total entropy change.  This yields a
fluctuation theorem for entropy changes in the steady
state \cite{Seifert05}
\begin{equation}
\frac{p(-\Delta \mathcal{S}_\text{tot})}{p(\Delta \mathcal{S}_\text{tot})}
=\rme^{-\Delta \mathcal{S}_\text{tot}}, \label{e:Sdet}
\end{equation}
which is essentially a stochastic form of the original fluctuation
theorem proposed by \citeasnoun{Evans93}.  

An analogous argument also
leads to~\eref{e:Sdet} for the case of periodic driving with an
integer number of symmetric cycles
\citeaffixed{Crooks99,Schuler05}{see e.g.,}.  Here $P^*$ means the
stationary distribution at the end of a complete period.

\subsection{Generalized symmetry relation}
\label{ss:gen}

In this subsection we show how the symmetry properties of
$\mathcal{R}_F$ and $\mathcal{R}_B$ lead also to a relationship for the
averages of other functionals of a trajectory.

Let us take a functional $\mathcal{X}_F$ of the type defined
by~\eref{e:X} together with the corresponding backward functional $\mathcal{X}_B$~\eref{e:XB}.
Since we now wish to consider functionals which are not necessarily
antisymmetric (i.e., in general we take
$r_{\sigma',\sigma}(\tau) \neq -r_{\sigma,\sigma'}(\tau)$) we also introduce the time-reversed functionals defined by,
\begin{eqnarray}
\mathcal{X}_{F,\text{rev}}(t,\{\sigma\},f,g) &\equiv
\mathcal{X}_{B}(t,\{\sigma\}_\text{rev},g,f) \\
\mathcal{X}_{B,\text{rev}}(t,\{\sigma\},g,f) &\equiv
\mathcal{X}_{F}(t,\{\sigma\}_\text{rev},f,g). 
\end{eqnarray}
As a concrete illustration, we note that for the example trajectory
introduced in figure~\ref{f:functional} we would have
$\mathcal{X}_{F,\text{rev}} = \ln g_2 +
r_{0,1}(\tau_1)+r_{1,2}(\tau_2) - \ln f_0$ and
$\mathcal{X}_{B,\text{rev}} = \ln f_2 + r_{0,1}(t-\tau_1) +
r_{1,2}(t-\tau_2) - \ln g_0$.  We remark that, in the special case of
antisymmetric $r$, we have $\mathcal{X}_F=-\mathcal{X}_{F,\text{rev}}$
and $\mathcal{X}_B=-\mathcal{X}_{B,\text{rev}}$.

To measure $\mathcal{X}_{F}$ we use the counting process with
increments $r_{\sigma',\sigma}(\tau)$.  It then follows
from~\eref{e:joint} that the joint generating function of
$\mathcal{X}_{F}$ and $\mathcal{R}_{F}$ under forward evolution is
given by
\begin{equation}
\langle \, \rme^{-\lambda \mathcal{R} - \lambda_\mathcal{X} \mathcal{X}}
\, \rangle^F
 = \langle \,s\, |\, P_B^{\lambda} g^{\lambda} 
T \{ \rme^{-\int_0^t
  \tilde{H}(\lambda,\lambda_\mathcal{X},\tau) \, \rmd \tau} \} 
f^{-\lambda} P_F^{-\lambda}\,|\, P_F \,\rangle.
\end{equation}
where $\tilde{H}(\lambda,\lambda_\mathcal{X},\tau)$ is the modified
Hamiltonian with off-diagonal elements
\begin{equation}
\left[\tilde{H}(\lambda,\lambda_\mathcal{X},\tau)\right]_{\sigma',\sigma} =
-w^F_{\sigma',\sigma}(\tau) \exp \left[-\lambda \ln
\frac{w^F_{\sigma', \sigma}(\tau)}{w^F_{\sigma, \sigma'}(\tau)} -
\lambda_\mathcal{X} r_{\sigma',\sigma} (\tau) \right]
\end{equation}
Now to measure the time-reversed backward functional
$\mathcal{X}_{B,\text{rev}}$ we use instead a counting process with
increments $r_{\sigma,\sigma'}(t-\tau)$.  The joint generating
function of $\mathcal{X}_{B,\text{rev}}$ and
$\mathcal{R}_{B,\text{rev}}$ under backward evolution is then given by
\begin{equation}
\langle \,\rme^{-\lambda \mathcal{R}_\text{rev} - \lambda_\mathcal{X} \mathcal{X}_\text{rev}}
\, \rangle^B
 = \langle \,s \,|\, P_F^{-\lambda} f^{-\lambda} 
T \{ \rme^{-\int_0^t
  \tilde{H}'(\lambda,\lambda_\mathcal{X},t-\tau) \, \rmd \tau} \} 
g^{\lambda} P_B^{\lambda}\,| \,P_B\, \rangle
\end{equation}
where $\tilde{H}'(\lambda,\lambda_\mathcal{X},\tau)$ is a subtly
different modified Hamiltonian with off-diagonal elements
\begin{equation}
\left[\tilde{H}'(\lambda,\lambda_\mathcal{X},\tau)\right]_{\sigma',\sigma} =
-w^F_{\sigma',\sigma}(\tau) \exp \left[-\lambda \ln
\frac{w^F_{\sigma, \sigma'}(\tau)}{w^F_{\sigma', \sigma}(\tau)} -
\lambda_\mathcal{X} r_{\sigma,\sigma'} (\tau) \right].
\end{equation}
Notice that for antisymmetric functionals (i.e.,
$r_{\sigma',\sigma}=-r_{\sigma,\sigma'}$) we have
$\tilde{H}'(\lambda,\lambda_\mathcal{X},\tau)=\tilde{H}(-\lambda,-\lambda_\mathcal{X},\tau)$.

In analogy to~\eref{e:Hsym2} one can prove the symmetry relation
\begin{equation}
\tilde{H}^T(\lambda,\lambda_\mathcal{X},\tau) 
= \tilde{H}'(\lambda-1,\lambda_\mathcal{X},t-\tau). \label{e:Hsymgen}
\end{equation}
and use it to transform the generating function:
\begin{eqnarray}
\fl \langle \, \rme^{-\lambda \mathcal{R} - \lambda_\mathcal{X} \mathcal{X}}
\, \rangle^F 
&= \langle\, s\, |\, P_B^{\lambda} g^{\lambda} 
T \{ \rme^{-\int_0^t
  \tilde{H}(\lambda,\lambda_\mathcal{X},\tau) \, \rmd \tau} \} 
f^{-\lambda} P_F^{-\lambda}\,|\, P_F \,\rangle \\
&= \langle\, P_F\, |\, P_F^{-\lambda}  f^{-\lambda} 
T \{ \rme^{-\int_0^t
  \tilde{H}^T(\lambda,\lambda_\mathcal{X},t-\tau) \, \rmd \tau}
\} 
g^{\lambda} P_B^{\lambda} \,|\, s\, \rangle \\
&= \langle \,s \,|\, P_F^{-(\lambda-1)}  f^{-\lambda} 
T \{ \rme^{-\int_0^t
  \tilde{H}'(\lambda-1,\lambda_\mathcal{X},t-\tau) \,
\rmd \tau} \} 
g^{\lambda} P_B^{\lambda-1} \,|\, P_B \,\rangle \\
&= \langle\, \rme^{-(\lambda-1) \mathcal{R}_\text{rev} 
- \lambda_\mathcal{X} \mathcal{X}_\text{rev}} \, \rangle^B. 
\end{eqnarray}
This provides a more general symmetry relation, viz.,  
\begin{equation}
\langle \, \rme^{-\lambda \mathcal{R} - \lambda_\mathcal{X} \mathcal{X}}
\, \rangle^F 
=\langle\, \rme^{-(\lambda-1) \mathcal{R}_\text{rev} 
- \lambda_\mathcal{X} \mathcal{X}_\text{rev}} \, \rangle^B. \label{e:symgen}
\end{equation}
In particular, if the functional $\mathcal{X}_F$ consists only of
boundary terms we recover~\eref{e:fundsym}.

\subsubsection{Crooks' relation}
Setting $\lambda=1$ in the ``generalized symmetry
relation''~\eref{e:symgen} leads to a statement about the average of
any function of $\mathcal{X}$
\begin{equation}
\langle \,A(\mathcal{X}) \rme^{-\mathcal{R}} \,\rangle^F = \langle \,A(\mathcal{X_\text{rev}})\, \rangle^B. \label{e:usymgen}
\end{equation}
This is reminiscent of a relation
derived by \citeasnoun{Crooks00} for discrete-time dynamics and
\emph{equilibrium initial conditions}.  
\begin{equation}
\langle \,\mathcal{F} \rme^{-\mathcal{R}} \rangle^F 
= \langle \mathcal{F}_\text{rev} \,\rangle^B. \label{e:Crooks}
\end{equation}
The Crooks' relation actually covers more general functionals of
trajectory, i.e., $\mathcal{F}$ is not restricted to $A(\mathcal{X})$
with $\mathcal{X}$ of the form~\eref{e:X}.  In fact, from~\eref{e:usymgen} with
$\mathcal{X}=\mathcal{R}$ 
one can recover all the results of
sections~\ref{ss:int}--\ref{ss:transient} \citeaffixed{Crooks00,Kurchan05}{for
examples of this approach see}.  In addition, we now have a
straightforward way to access some other relationships as explored
below.

\subsubsection{Symmetry relations for currents}

If $\mathcal{X}$ is a pure current $\mathcal{J}_r$ 
(i.e., no boundary terms and increments obeying
$r_{\sigma',\sigma}=-r_{\sigma,\sigma'}$, see
section~\ref{ss:functionals}), then by antisymmetry~\eref{e:usymgen}
gives
\begin{equation}
\langle\, A(\mathcal{J}_r)\, \rme^{-\mathcal{R}} \rangle^F 
= \langle\, A(-\mathcal{J}_r) \,\rangle^B.   \label{e:Jsym}
\end{equation}
Note that for any time-independent equilibrium process with
$P_F=P_B=P^\text{eq}$, then
$\mathcal{R}_F=\mathcal{R}_B=\mathcal{W}_d=\Delta\mathcal{S}=0$ by detailed balance
and~\eref{e:Jsym} becomes an (obvious) statement about equilibrium
current fluctuations.  

In section~\ref{s:asymp} we give a detailed discussion of symmetry
relations for currents in the long-time limit.

\subsubsection{Kawasaki response and related relations}

With the choice $A(\mathcal{X}) = h(\sigma(t))$ (where $h$ is an arbitrary
function), equation~\eref{e:usymgen} gives
\begin{equation}
\langle \, h(\sigma(t)) \rme^{-\mathcal{R}} \,\rangle^F 
= \langle\, h(\sigma(0)) \,\rangle^B. \label{e:intgen}
\end{equation}
Notice that in the reverse process we have the average of a function
of the initial condition.  Since this is an average over the
distribution $P_B(\sigma)$, the (reversed) protocol
plays no role.  Note that~\eref{e:intgen} also follows directly from the
$\lambda=1$ ``integral'' version of~\eref{e:symgen} with $X=0$. Just
as in section~\ref{ss:int}, the two most physically relevant cases for
the boundary distributions $P_F$ and $P_B$ are
\begin{itemize}

\item $P_F=P^\text{eq}_{l^F(0)}$ and $P_B=P^\text{eq}_{l^F(t)}$, giving
\begin{equation}
\langle \, h(\sigma(t)) \rme^{-\mathcal{W}_d}\, \rangle^F
 = \langle \,h(\sigma(0)) \,\rangle^B \\
\end{equation}
which we can write as
\begin{equation}
\langle \, h(\sigma(t)) \rme^{-\mathcal{W}_d}\, \rangle_{\text{neq}} 
= \langle \, h(\sigma(t))\, \rangle_{\text{eq}, l(t)} \\ \label{e:Kawasaki}
\end{equation}
to highlight that we now have a relation between a non-equilibrium
average and an equilibrium average of the same function.  This
equality, originally due to Jarzynski, is closely related to the
Kawasaki nonlinear response relation \cite{Yamada67};
see \citeasnoun{Crooks00} for a discussion.

\item $|\,P_B\,\rangle = T \{ \rme^{-{\int_0^t H(\tau)}} \}\,|\,P_F\,\rangle$, giving
\begin{eqnarray}
\langle \, h(\sigma(t)) \rme^{-\Delta \mathcal{S}}\, \rangle^F 
&= \langle\, h(\sigma(0))\, \rangle^B \\
&= \langle\, h(\sigma(t))\, \rangle^F.
\end{eqnarray}
This relationship was derived by \citeasnoun{Schmiedl06} directly
from~\eref{e:Rint} with the choice
\begin{equation}
|\,P_B\,\rangle = \frac{h  \, 
T \{ \rme^{-{\int_0^t H(\tau)}} \}\,|\,P_F\,\rangle }{\langle \, h(\sigma(t)) \,\rangle^F}.
\end{equation}
where $h$ is a diagonal matrix with elements $h(\sigma)$.
\end{itemize}

\section{Asymptotic fluctuation relations}
\label{s:asymp}

\subsection{Fluctuations in the long-time limit}

In this section we focus on the case of \emph{time-independent} rates
and discuss a class of fluctuation relationships which are expected to
hold asymptotically for $t \to \infty$.  In this long-time limit an
ergodic system has always relaxed into a unique stationary state,
hence these fluctuation theorems can also be described as
``steady-state''.  However, we emphasize that our discussion does not
not require the system to have relaxed to stationarity before the
start of the measurement; in general we will consider an arbitrary
initial condition $P_0$.

Firstly, we remark that all the transient relations of
section~\ref{s:finite} hold (of course) also in the limit $t \to
\infty$.  In particular, recall from section~\ref{sss:ESFT}, that choosing $P_F=P_B=P_0$
in the functional $\mathcal{R}$~\eref{e:Rdef} leads to the Evans-Searles Fluctuation
Theorem
\begin{equation} 
\frac{p(-{\Omega})}{p({\Omega})}=\rme^{-{\Omega}}. \label{e:ESR}
\end{equation}
for the dissipation function,
\begin{equation}
{\Omega}(t,\{\sigma\})=\mathcal{J}_{\mathcal{Q}}(t,\{\sigma\})
-\ln \frac{P_0(\sigma(t))}{P_0(\sigma(0))} \label{e:diss}.
\end{equation}
When the initial distribution $P_0$ is identical to the stationary
distribution $P^*$, $\Omega$ can be identified with the total entropy
change (see section~\ref{sss:entropy}).

Now notice that the first term in~\eref{e:diss} is an integrated
current which, in the steady state, grows on average linearly with
$t$.  One then naively expects that, in the long-time limit, the
contribution of the initial condition terms in~\eref{e:diss} can be
neglected.  In that case, in addition to~\eref{e:ESR} one has the
asymptotic relation
\begin{equation} 
\frac{p(-\mathcal{J}_{\mathcal{Q}})}{p(\mathcal{J}_{\mathcal{Q}})} \sim
\rme^{-\mathcal{J}_{\mathcal{Q}}}. \label{e:GCFT}
\end{equation} 
Recall that $\mathcal{J}_{\mathcal{Q}}$ can be interpreted as the
entropy production of the environment (which differs by boundary terms
from the total entropy change).  Hence this equation is equivalent to the
Gallavotti-Cohen Fluctuation Theorem \cite{Gallavotti95,Gallavotti95b}
for the steady state of deterministic systems (where the entropy
production is identified with the phase-space contraction rate). In
fact, in the following subsection we will show that a relationship of
the form~\eref{e:GCFT} is also expected for other currents weighted by
appropriate ``conjugate fields''.  This leads to a symmetry property
of the large deviation function which we will refer to as
Gallavotti-Cohen symmetry (GC symmetry).

As we shall see below, the argument of the preceding paragraph can be
made rigorous only for the case of finite state space.  It has
recently been recognised that for infinite state space (or unbounded
potentials) there can be a breakdown in the Gallavotti-Cohen symmetry,
see e.g., \citeasnoun{VanZon03}, \citeasnoun{Bonetto06b},
\citeasnoun{Me06b}.  Essentially, this arises from the fact that the
boundary probabilities appearing in~\eref{e:diss} can be arbitrarily
small.  In section~\ref{ss:break}, we explore this issue further and
discuss some examples.

\subsection{Gallavotti-Cohen symmetry for currents}
\label{ss:GC}

The first derivation of~\eref{e:GCFT} for general Markov processes
with bounded state space was given by \citeasnoun{Lebowitz99}.  A concise
proof of the GC symmetry for \emph{particle currents} in a specific
model can also be
found in \citeasnoun{Derrida04b}.  In the following we generalize this
approach to discuss \emph{arbitrary currents} of the
form~\eref{e:Jr} \cite{MeARunpub}. Our abstract discussion is
  complemented by concrete treatment for a toy
model in
\ref{s:app}.  As an aside, we remark that equivalent
forms of current fluctuation theorem have also been obtained using
Schnakenberg network theory \cite{Andrieux05}.

We start with an equilibrium system with rates
$w^\text{eq}_{\sigma',\sigma}$ obeying detailed balance.  Now we
imagine obtaining another model by applying a field $E_r$ 
conjugated to some current
$\mathcal{J}_r$ (defined by $r_{\sigma',\sigma} \equiv
-r_{\sigma,\sigma'}$).  Specifically this means that we have a new process defined by
transition rates 
\begin{equation}
w_{\sigma',\sigma}=w^\text{eq}_{\sigma',\sigma} \rme^{\frac{E_r}{2}
r_{\sigma',\sigma}}, \label{e:eq}
\end{equation} 
with (by construction)
\begin{equation}
\frac{w_{\sigma',\sigma}}{w_{\sigma,\sigma'}}
=\frac{w^\text{eq}_{\sigma',\sigma}}{w^\text{eq}_{\sigma,\sigma'}}
\rme^{E_r r_{\sigma',\sigma}}. \label{e:localdb}
\end{equation} 
In general, the rates $w_{\sigma',\sigma}$ define a non-equilibrium
model without detailed balance.  In this case, the fluctuations of the quantity $E_r
\mathcal{J}_r$ in the ``driven'' system obey the Gallavotti-Cohen
symmetry, as we shall prove below.

Note that for a given non-equilibrium model with rates
$w_{\sigma',\sigma}$, there may be several different currents whose
conjugate fields obey~\eref{e:localdb} with \emph{different} rates for
the corresponding ``undriven'' process.  In the quantum Hamiltonian
formalism introduced in section~\ref{s:formalism}, we can always
construct another stochastic matrix from $H$ by replacing the
transition rates $w_{\sigma', \sigma}$ with $w_{\sigma',
\sigma}\rme^{-\lambda r_{\sigma', \sigma}}$ in \emph{both} diagonal
and off-diagonal elements.  If for some specific $\lambda$ this new
matrix obeys the detailed balance condition~\eref{2-7} (with
corresponding equilibrium distribution $P_r^\text{eq}$) then this
value is the required conjugate field $E_r/2$.  In other words, the
statement for a general non-equilibrium system is: \emph{if} for a
given current $J_r$ there exists a field $E_r$ such that the rates
satisfy
\begin{equation}
w_{\sigma',\sigma}\rme^{-\frac{E_r}{2}r_{\sigma',\sigma}} \times
P_r^\text{eq}(\sigma) =
w_{\sigma,\sigma'}\rme^{-\frac{E_r}{2}r_{\sigma,\sigma'}} \times
P_r^\text{eq}(\sigma'), \label{e:localdb2}
\end{equation}
\emph{then} the fluctuations of $E_r
\mathcal{J}_r$ obey the Gallavotti-Cohen symmetry. 
We remark that, for the particular case where
$\mathcal{J}_r$ is the total integrated particle current in a lattice gas and $E_r$ a
constant external field, the condition \eref{e:localdb} (or
equivalently~\eref{e:localdb2}), is known as
``local detailed balance'' \cite{Katz84} and leads to the GC symmetry
for fluctuations of particle current \cite{Lebowitz99}.

In order to demonstrate the symmetry for general
$\mathcal{J}_r$, we begin by considering the long-time limit of the current 
generating function, viz.,
\begin{equation} 
e_r(\lambda) =\lim_{t \to \infty} - \frac{1}{t} \ln
{\langle \,\rme^{-\lambda \mathcal{J}_r(t)} \,\rangle}. \label{e:e_phi}
\end{equation} 
As observed in \citeasnoun{Lebowitz99}, the existence of this limit implies
a large deviation property for the probability distribution,
$p(j_r,t)$, of the observed ``average'' current
$j_r=\mathcal{J}_r/t$.  [Here we write the time-dependence explicitly
to remind the reader that although we have divided the integrated
current by the elapsed time the distribution of the average can still
be time-dependent].  Specifically, the long-time limiting behaviour is
given by
\begin{equation} 
p(j_r,t) \sim \rme^{-t\hat{e}_r(j_r)} \label{e:pj}
\end{equation} 
where the large deviation function $\hat{e}_r(j)$ is the Legendre
transformation of $e_r(\lambda)$, i.e.,\footnote{This can be shown by
performing a Fourier transform on the generating function and then
using a saddle-point approximation to evaluate the resulting integral
for large times.  A mathematically rigorous treatment exists providing
$e(\lambda)$ is differentiable; see, e.g., \cite{Ellis85,Dembo98}.}
\begin{equation} 
\hat{e}_r(j)=\max_{\lambda}\{e_r(\lambda)-\lambda j\}. \label{e:lang}
\end{equation} 

Now recall that we can write the generating function in terms of a
modified Hamiltonian $\tilde{H}_r$ with off-diagonal elements
$-w_{\sigma',\sigma} e^{-\lambda r_{\sigma',\sigma}}$.  The next step
is to insert a complete set of eigenstates of this operator:
\begin{eqnarray} 
\langle \, s\, | \, e^{-\lambda \mathcal{J}_r} \,|\, P_0\, \rangle
&= \sum_k \langle \,s \,|\, e^{-\tilde{H}_r(\lambda)t}\, |\, P_0 \,\rangle \\ 
&= \sum_k \langle \,s \,|\, \psi_{k}(\lambda) \,\rangle e^{-\epsilon_{k}(\lambda) t} \langle\,
\psi_{k}(\lambda) \,|\, P_0 \,\rangle \label{e:insert}
\end{eqnarray} 
where $|\,\psi_{k} (\lambda)\, \rangle$ are the eigenvectors of
$\tilde{H_r}$ and $\epsilon_{k}(\lambda)$ the corresponding
eigenvalues (for notational simplicity we suppress the $r$-dependence
of these quantities).  For bounded state space, all the products
$\langle\, s\, | \,\psi_k \,\rangle$ and $\langle\, \psi_k \,| \,P_0\, \rangle$ are
finite and it follows that in the long-time limit we can identify
$e_r(\lambda)$ with the lowest eigenvalue $\epsilon_{0}(\lambda)$.
Note that in complete analogy to equilibrium, the Legendre
transformation relates nonextensive and extensive variables. 
The logarithm $t e_r(\lambda)$ of the generating function for
the integrated current
is extensive in time (and in spatially extended particle
systems generically also in the volume). This extensivity is
analogous to the role of volume $V$ in equilibrium 
large deviation theory (cf.\
section~\ref{s:intro}). 

Now, for $\mathcal{J}_r$ with conjugate field $E_r$
obeying~\eref{e:localdb2}, one can easily demonstrate the symmetry
relation
\begin{equation} 
\tilde{H}_r^T(\lambda)
=(P^{\text{eq}}_r)^{-1} \tilde{H}_r(E_r-\lambda) P_r^{\text{eq}}, \label{e:GCFTH}
\end{equation} 
where $P^{\text{eq}}_r$ is the diagonal matrix with elements which are
the equilibrium probabilities of the corresponding undriven system.  Note that for
the case $r_{\sigma',\sigma}=\ln [w_{\sigma',\sigma}/
w_{\sigma,\sigma'}]$ (i.e.,
$\mathcal{J}_r=\mathcal{J}_{\mathcal{Q}}$), then~\eref{e:localdb2} is
satisfied with conjugate field $E=1$ and equal equilibrium probabilities;
the symmetry relationship
\eref{e:GCFTH} then reduces to (\ref{e:Hsym2}).

Significantly,~(\ref{e:GCFTH}) means
that the eigenvalues of $\tilde{H}_r(\lambda)$ and
$\tilde{H}_r(E_r-\lambda)$ are identical and hence imposes the
Gallavotti-Cohen symmetry in the form
\begin{equation} 
e_{r}(\lambda)= e_{r}(E_r-\lambda). \label{e:GCFTe}
\end{equation} 
This leads via~\eref{e:pj} and~\eref{e:lang} to the more usual form
\begin{equation} 
\frac{p(-j_r,t)}{p(j_r,t)} \sim \rme^{-E_r j_r t}
\label{e:GCFTj}
\end{equation} 
which is trivially equivalent to
\begin{equation} 
\frac{p(-\mathcal{J}_r)}{p(\mathcal{J}_r)} \sim \rme^{-E_r \mathcal{J}_r} 
\label{e:GCFTJ}.
\end{equation}
 
In other words, in the asymptotic limit we obtain a fluctuation
theorem for $E_r \mathcal{J}_r$ (i.e., \emph{any} current for which a
conjugate field obeying~\eref{e:localdb}/\eref{e:localdb2} exists) for arbitrary initial
distribution $P_0$.  This is a slightly stronger statement than that
originally obtained by \citeasnoun{Lebowitz99}.  It also
follows that this asymptotic theorem holds if we have a protocol which
is time-dependent from $\tau=0$ to some finite time $\tau_c$ and then
time-independent from $\tau_c$ to $t$ (this is equivalent to altering
the initial distribution).  In fact, a similar argument can be used to
show that the large deviation function is unchanged by any finite
section of time-dependent driving (i.e., one that forms a vanishing
fraction of the total history).

One can explicitly show \cite{MeARunpub} that any two currents which
satisfy the conditions of the fluctuation theorem differ only in
boundary terms.  Specifically,
\begin{equation} E_r \mathcal{J}_r =\mathcal{J}_{\mathcal{Q}}
-\ln\frac{P_r^\text{eq}(\sigma(t))}{P_r^\text{eq}(\sigma(0))}.
\end{equation}
Notice also, that if $P_0=P_r^\text{eq}$ the boundary terms cancel
with those in the definition of $\Omega$~\eref{e:diss} and we have
$\Omega=E_r \mathcal{J}_r$.  Indeed for arbitrary $P_0$ and
time-independent protocols, this gives a physical interpretation of
$\Omega$---it is (proportional to) the heat dissipated by that
particular current for which the initial distribution would be the
undriven equilibrium state.  

Calculating the large deviation function for a given current in a
stochastic model is typically a difficult problem.  However there has
recently been notable success using both microscopic and macroscopic
approaches \cite{Derrida04b,Bodineau06,Bertini06b}.
For example, in \citeasnoun{Derrida04b}, the first four cumulants of the
distribution of particle current from a boundary reservoir into a
symmetric exclusion process are calculated and explicitly shown to
obey the Gallavotti-Cohen symmetry with appropriate field.  

We conclude this subsection by remarking that one can readily extend
the analysis to the joint probability distributions of two different
currents.  For example, it is already clear from~\eref{e:Hsymgen} that
for the joint probability of $\mathcal{J}_{\mathcal{Q}}$ and any other antisymmetric
current $\mathcal{J}$, the large deviation function has the symmetry
\begin{equation}
e(\lambda,\lambda_r)=e(1-\lambda,-\lambda_r)
\end{equation} 
Another interesting result is obtained by counting separately the
current between each pair of states for which $w_{\sigma',\sigma}>0$,
i.e, we set $r^{(i)}_{\sigma',\sigma}=\delta_{\sigma,\sigma_i}
\delta_{\sigma',\sigma_{i'}} - \delta_{\sigma,\sigma_{i'}}
\delta_{\sigma',\sigma_i}$ for each of the $m$ such pairs.  In this
case, the analogue to~\eref{e:GCFTe} is
\begin{equation} 
e(\lambda_1, \ldots, \lambda_m) 
= e(E_1-\lambda_1,\ldots, E_m-\lambda_m) \label{e:ejoint}.
\end{equation} 
Note also that, in the equilibrium limit, the Gallavotti-Cohen
symmetry implies the Green-Kubo formula and Onsager reciprocity
relations \cite{Gallavotti96}.  \Eref{e:ejoint} is the starting point
for a proof within the present stochastic
framework---see \citeasnoun{Lebowitz99}.  A recent discussion of the
implications of the GC symmetry for \emph{non-linear} response
coefficients can be found in \citeasnoun{Andrieux07}.

\subsection{Breakdown for unbounded state space}
\label{ss:break}

The importance of boundary terms in distinguishing transient and
asymptotic fluctuation theorems has been recognised for some
time \cite{Evans02b}.  However, it is only relatively recently that
the dramatic effects of such boundary terms in the case of unbounded
state space have been elucidated.  Not only do the large deviation
functions now depend on the initial condition, but a breakdown of the
asymptotic symmetry relation~\eref{e:GCFT} can result.

In the deterministic context, the effect of unbounded potentials was
discussed by \citeasnoun{Bonetto06b}.  They argued that the violation
of the fluctuation theorem in some simulation results (see
\citeasnoun{Evans05c} and references therein) was due to the
possibility of arbitrarily large potential energies and proposed the
restoration of the symmetry by removal of the ``unphysical'' singular
terms; see also \citeasnoun{Zamponi07}.  An earlier study of a
particular dissipative model \cite{VanZon03,VanZon04} attributed a
modified form of heat fluctuation theorem to a combination of
deterministic and stochastic forces treated via a Langevin
equation.  
Related discussions for Langevin dynamics can also be found in
\citeasnoun{Farago02}, \citeasnoun{Baiesi06}, \citeasnoun{Visco06}.
In the present subsection we review the significance of unbounded
state space for many-particle stochastic dynamics, see
\citeasnoun{Me06b}, \citeasnoun{Puglisi06}, \citeasnoun{MeARunpub}.

The derivation of section~\ref{ss:GC} only applies in the case of
finite state space.  Specifically, for infinite state space a problem
arises where we inserted a complete set of eigenstates.  For unbounded
state space the scalar products appearing in~\eref{e:insert} are not
guaranteed to be finite and one cannot simply identify the large
deviation function with the lowest eigenvalue.  To be more exact,
$\langle \,s\, |\, \psi_k\, \rangle$ and $\langle \,\psi_k \,|\, P_0\, \rangle$ become
infinite series with a (possibly finite) regime of convergence in
$\lambda$-space.  In general, the radius of convergence depends both
on the particular current studied (through the form of the
eigenvectors) and, for $\langle \,\psi_k \,| \,P_0\, \rangle$, also on the
initial distribution $P_0$.

A concrete example is provided by \citeasnoun{Me06b} in which current
fluctuations are studied in a simple stochastic particle system---the
zero-range process with open boundaries \cite{Levine04c}.  In this
model, the number of particles on each lattice site is unlimited so
that, even for a finite lattice one has an infinite state space.  For
the particle current across a particular bond, the lowest eigenstate of
$\tilde{H}(\lambda)$ can be explicitly calculated \cite{Me05} leading
to expressions for the corresponding radii of convergence and hence a
(bond-dependent) restricted range of validity of the
symmetry~\eref{e:GCFTj}.  Physically, the breakdown in GC symmetry for
large current fluctuations results from ``instantaneous
condensation''---the temporary piling-up of an arbitrarily large number of
particles on a site.  For a single-site model, exact calculations for
the initial-condition-dependent behaviour of $p(-j,t)/p(j,t)$ outside
the GC-symmetry regime show good agreement with simulation.  In contrast
to numerical and analytical treatments for other models (see e.g.,
\citeasnoun{VanZon03}, \citeasnoun{Bonetto06b}, \citeasnoun{Baiesi06},
\citeasnoun{Puglisi06}), a constant value for the ratio of
probabilities for large forward and backward fluctuations is not
found.  It is argued in \cite{MeARunpub} that the failure to observe
this form of ``extended fluctuation theorem'' is due to
correlations between the boundary terms and the integrated current
which persist even in the long-time limit. 

\subsection{Periodic driving}

We now conclude our treatment of asymptotic fluctuation theorems by
considering how to apply the above arguments to the experimentally
relevant case of a time-periodic protocol.  

First we remark that one can always define an effective
time-independent Hamiltonian for a complete cycle
(period $\Delta t$) via the following relation,
\begin{equation} 
\rme^{-H^\text{eff} \Delta t} 
= T\{\rme^{-\int_0^{\Delta t} H(\tau) \, d\tau} \}.
\end{equation} 
Hence we can treat the case of an integer number
of cycles equivalently to the time-independent case with Hamiltonian
$H^\text{eff}$.

Now for a time-symmetric cycle, since
$\tilde{H}_\mathcal{Q}(\lambda,\tau)$ obeys the
symmetry~\eref{e:Hsym2} for $0<\tau<\Delta t$, it follows that
$\tilde{H}^\text{eff}_\mathcal{Q}(\lambda)$ also has this symmetry.
Hence, \emph{for bounded state space}, we expect that
$\mathcal{J}_{\mathcal{Q}}$ will obey the Gallavotti-Cohen FT~\eref{e:GCFT} in
the limit of a large number of complete cycles (the finite
initial-condition-dependent boundary terms can be neglected).
Furthermore for increasing times, we expect the violation from the
symmetry for a non-integer number of cycles to become smaller and
smaller (cf.\ the argument following~\eref{e:GCFTJ}).  However, in
contrast to the time-independent case, we do not in general expect a
fluctuation relation for other currents.  Formally, this is because
although the modified Hamiltonian for such a current
$\tilde{H}_r(\lambda,\tau)$ obeys the symmetry relation~\eref{e:GCFTH}
for $0<\tau<\Delta t$ [with time-dependent $E_r(\tau)$ and
$P^\text{eq}_r(\tau)$], it does not necessarily follow that the
effective Hamiltonian for a complete cycle
$\tilde{H}^\text{eff}_r(\lambda,\tau)$ has a similar symmetry.
Physically, this point is well-illustrated by thinking of cases where
the effective driving fields for different currents (e.g., across
different bonds in an exclusion process) are out of phase.

In the case of \emph{unbounded state space}, we anticipate the
boundary terms arising from the initial condition and the finite
non-symmetric part, to be relevant in determining the symmetries of
$\mathcal{J}_{\mathcal{Q}}$.  Indeed a breakdown in the Gallavotti-Cohen FT
for heat fluctuations in a periodically driven oscillator was
recently found experimentally and supported by a second-order Langevin
analysis \cite{Joubaud06}. 

\section{Simulation and experiment}
\label{s:exp}

Numerical simulations of sheared fluids \cite{Evans93} were the
original motivation for the development of fluctuation theorems.
Since the experiments of \citeasnoun{Ciliberto98} on turbulent
Rayleigh-Benard convection there have been many attempts at experimental
verification.  Such empirical studies are often plagued by technical
issues and problems of interpretation.  For example, to test
steady-state/asymptotic theorems one desires to measure fluctuations
over long times but, as time is increased, it becomes exponentially
more unlikely to see fluctuations away from the mean.  On the other
hand, verification of transient theorems relies on the correct
identification and accurate measurement of thermodynamic quantities
such as work and entropy/heat.  Despite these caveats, there have by
now been successful experimental/numerical tests for several different
types of system and even some cases in which useful predictions can be
made (notably, the extraction of free energy differences from work
distributions).  In the following subsections, we provide a few
pointers to the relevant literature whilst making no attempt at
completeness.  For more details, the interested reader is referred to
the reviews by \citeasnoun{Ritort03} and
\citeasnoun{Evans02b}. Further discussions, for 
Langevin and deterministic dynamics respectively, can be found in
\citeasnoun{Narayan04} and \citeasnoun{Zamponi07}.

\subsection{Tests and predictions}
\label{ss:test}

Direct testing of fluctuation theorems typically requires one to look for a
relationship of the form
\begin{equation}
\frac{p(-\mathcal{X})}{p(\mathcal{X})} = \rme^{-\mathcal{X}} 
\label{e:Rsimp}
\end{equation}
where for simplicity we here restrict the discussion to the case of a
time-symmetric protocol.  Depending on the exact definition of
$\mathcal{X}$,~\eref{e:Rsimp} is expected to told at finite-times (see
section~\ref{s:finite}) or just in the long-time limit (see
section~\ref{s:asymp}).  The obvious way to check for such a form is to
build up a distribution histogram by coarse-graining the data into
``bins'' so that $N(\mathcal{X})$ represents the number of
experimental realizations with a measured value of $\mathcal{X}$
between $\mathcal{X}- \Delta_\mathcal{X}/2$ and $\mathcal{X}+
\Delta_\mathcal{X} /2$. One then plots $-\ln [N(-\mathcal{X}) /
N(\mathcal{X}) ]$ versus $\mathcal{X}$; if~\eref{e:Rsimp} holds then a
straight line of slope 1 should result.  However, sometimes the
quantity and quality of experimental data make such a procedure
difficult; in particular there are often very few realizations with
negative values of $\mathcal{X}$.  Hence, as detailed below, various
indirect tests have been developed which check \emph{predictions} of
the FT rather than the FT itself.

\begin{itemize}

\item If $\mathcal{X}$ is assumed to have a Gaussian distribution
(typically this is only expected to be a good approximation close to
equilibrium, although it may also hold further away) with mean
$\bar{\mathcal{X}}$ and standard deviation $\sigma_\mathcal{X}$ then
consistency with~\eref{e:Rsimp} requires
$\sigma_\mathcal{X}=\sqrt{2\bar{\mathcal{X}}}$.  Such a relation was
found to hold approximately in early experiments with liquid-crystal
electroconvection \cite{Goldberg01}.

\item A straightforward consequence of~\eref{e:Rsimp} is the
relationship
\begin{equation} 
\langle e^{-\mathcal{X}} \rangle_{\mathcal{X}>a} =
\frac{\sum_{\mathcal{X}<-a} p(\mathcal{X})}{\sum_{\mathcal{X}>a}
p(\mathcal{X})}
\end{equation} 
(formally this follows from~\eref{e:usym} by inserting a step function
of $\mathcal{X}$).  Setting $a=0$ gives the ``integrated fluctuation
relation'' \cite{Ayton01}---in this case the right hand side takes the
form ``number of entropy decreasing trajectories/number of entropy
increasing trajectories''.  This form is used, for example, to test
the experimental data in \citeasnoun{Wang02} and
\citeasnoun{Carberry04}.  In the first of these studies, anomalous
results for small times were attributed to difficulties in
distinguishing between small positive and small negative values of
$\mathcal{X}$; in simulations, the present authors have found that
taking a small positive $a$ reduces numerical errors.

\item The following relation for the moments of the distribution also
follows from~\eref{e:usym}
\begin{equation} 
\langle \,(\mathcal{X})^k e^{-\mathcal{X}} \,\rangle
= (-1)^k \langle \,(\mathcal{X})^k \,\rangle. \label{e:moment}
\end{equation} 
For example, in \citeasnoun{Schuler05} the FT is tested by checking this
relationship for the second and fourth moments.  Of course, checking
an infinite number of moments becomes equivalent to a direct test of
the original FT.

\end{itemize}

We conclude this subsection with a few remarks relating specifically
to the experimental study of the Crooks fluctuation
theorem~\eref{e:CrooksFT}.  Since $\mathcal{W}_d=\mathcal{W}-\Delta F$
this can alternatively be written as
\begin{equation}
\frac{p_B(\mathcal{-W})}{p_F(\mathcal{W})}=\rme^{-(\mathcal{W}-\Delta F)/T}
\label{e:CrooksFTb}.
\end{equation} 
It now follows that the
distributions of $p_F(\mathcal{W})$ and $p_B(-\mathcal{W})$ should
cross at the value $\mathcal{W}=\Delta F$.  This provides a simple way
to test the theorem or, more significantly, to predict an unknown free
energy difference from the distribution of work in forward and reverse
processes.  However, this
method does not provide a very reliable estimate of $\Delta F$ since
it uses only a small part of the distribution.  There is a vast body
of literature discussing how to improve statistics and minimize errors
for free energy estimation
using the Crooks FT or the related Jarzynski
equality~\eref{e:Jar}, see e.g., \citeasnoun{Zuckerman02},
\citeasnoun{Gore03}, \citeasnoun{Shirts03} and \citeasnoun{Schmiedl07}.

\subsection{Numerical techniques}
\label{ss:sim}

Direct simulation for stochastic systems typically involves some form
of Monte Carlo algorithm; in the context of fluctuation theorem tests
for simple models, see e.g, \citeasnoun{Crooks99}, \citeasnoun{Me06b},
\citeasnoun{Puglisi06}.  Compared to experiments, one has the considerable advantage of
precise knowledge and control of the initial distribution but the
problem of exponentially small probabilities still remains.  One way
to get round this difficulty is by weighting the dynamics so that rare
events become typical.  An efficient algorithm using such a method to
calculate large deviation functions was introduced by
\citeasnoun{Giardina06} and tested on both
stochastic and deterministic systems (the latter case including an
explicit check of the Gallavotti-Cohen fluctuation theorem for a
Lorentz gas).  Continuous time versions of the algorithm have been used
for the symmetric exclusion process \cite{Lecomte06c} and the
zero-range process \cite{MeARunpub}.  Many computer simulation studies
are motivated by experiment---see the next subsection for some
examples.

\subsection{Experimental studies}
\label{ss:exp}

We here list a few different classes of systems in which fluctuation
theorems have been investigated experimentally.  In particular, we
emphasize that the final two items contain situations which are
naturally interpreted within a stochastic master equation approach.

\begin{itemize}

\item A number of works have studied colloidal particles in
time-dependent traps, yielding distributions of heat and work
consistent with fluctuation theorem predictions.  However, the first
experiments in this direction \cite{Wang02,Carberry04} had harmonic
potentials which is a rather special case (distributions of relevant
quantities are usually Gaussian) see, e.g., the discussion in
\citeasnoun{Baiesi06}.  A more recent study treated the case of a
non-harmonic potential and found good agreement with both the
Jarzynski relation and a work fluctuation theorem \cite{Blickle06}.
Other works include comparison of transient and steady-state
fluctuation theorems \cite{Wang05c} and investigation of entropy and
heat for individual trajectories \cite{Andrieux07b}.  In all these
experiments the motion of the colloidal particle can be described by a
Langevin equation, \citeaffixed{vanZon03b}{see e.g.,}.

\item An analogous equation to the Langevin equation for a colloid
appears in an analysis of non-equilibrium fluctuations in electric
circuits \cite{vanZon04b}.  Experimental work \cite{Garnier05}
includes an indication of the breakdown of the Gallavotti-Cohen
symmetry for heat fluctuations, in agreement with theoretical
predictions \cite{VanZon03,VanZon04}.  Furthermore, it is suggested
in \citeasnoun{Garnier05} that the symmetry of fluctuations can actually be
used to predict an unknown averaged dissipated power.

\item Single molecule manipulation experiments provide a particularly
fertile ground for testing and applying the FT by
Crooks~\eref{e:CrooksFT} and the Jarzynski equality~\eref{e:Jar}---see
\citeasnoun{Jarzynski06c} for an overview.  Typical scenarios involve
pulling molecules with laser optical tweezers or an atomic force
microscope, for example to mechanically fold and unfold a protein
\cite{Hummer01,Liphardt02,Collin05}.  Such experiments have
successfully demonstrated the recovery of equilibrium free energies
under strong non-equilibrium conditions.  \citeasnoun{Ritort06b}
contains a comprehensive review of single-molecule experiments,
including a section on testing fluctuation/work theorems.  Related
theoretical work includes \citeasnoun{Ritort02}, \citeasnoun{Braun04},
\citeasnoun{Imparato06} and \citeasnoun{Imparato06c}.

\item Fluctuations in granular gases have been studied via both
simulation \cite{Aumaitre01} and experiment \cite{Feitosa04}.  The
standard model of such granular media is a gas of inelastic hard
spheres but Visco, Puglisi and co-workers argue that one cannot use
this model to test fluctuation theorems since individual collisions
violate time reversibility and the time-reversed trajectory does not
exist \cite{Visco06b,Puglisi06}.  However, the dynamics of a single
granular tracer particle can be formulated in terms of a continuous
time Markov process \cite{Puglisi06b}.  Numerical studies of such
systems provide an illuminating example of the breakdown of the
Gallavotti-Cohen symmetry \cite{Puglisi06b,Puglisi06}.  An alternative
deterministic approach can be found in \citeasnoun{Bonetto06}.

\item The optical excitation of a single-defect centre in diamond has
been studied as a simple realization of a two-level system with
time-dependent rates.  Measurements of the dissipated work
\cite{Schuler05} and entropy \cite{Tietz06b} agree with the expected
integral fluctuation relations together with stronger fluctuation
theorems (generating-function symmetries) for an integer number of
driving periods. (The data also suggest that, as expected, the
deviation for a non-integer number of periods becomes smaller with
increasing time.)  Results from numerical solution of the
Chapman-Kolmogorov equation are presented in \citeasnoun{Schuler05}
whereas the same system is treated in \citeasnoun{Imparato06d} by an
algorithm which generates trajectories with judiciously chosen
statistical weights (cf.\ the discussion of the previous subsection);
both give good agreement with the experimental results.  Finally, we
remark that chemical reaction networks have been suggested as a more
general testbed for multi-state processes \cite{Schmiedl06b}.

\end{itemize}

\section{Final remarks}
\label{s:dis}

The experimental and theoretical study of fluctuation theorems has
been a major field of activity in non-equilibrium physics in the last
decade and is likely to continue to be fruitful in coming years.  From
a theoretical perspective, we point out that, so far, the focus of
attention has been on time-reversal relations for the functional
\eref{e:Rdef}. However, other functionals can also be studied. We
mention here some possibilities.

(i) Instead of comparing the probability of a trajectory
with its time-reversed image under the backward evolution we can define
a functional that arises from the probability ratio of a trajectory to the
same trajectory under the adjoint time evolution defined in Section~\ref{sss:ser}.  
This leads us to consider the functional\footnote{Note that here we do not
explicitly indicate a dependence on the direction of the
protocol, since we will have no need to compare forward and backward
functionals in the analysis.}
\begin{equation}
\mathcal{T}(t,\{\sigma\}) =\ln \left[
\frac{w_{\sigma_n,\sigma_{n-1}}(\tau_n) \ldots
w_{\sigma_2,\sigma_1}(\tau_2) \, w_{\sigma_1,\sigma_0}(\tau_1)}
{w^\text{ad}_{\sigma_n,\sigma_{n-1}}(\tau_n) \ldots
w^\text{ad}_{\sigma_2,\sigma_1}(\tau_2) \,
w^\text{ad}_{\sigma_1,\sigma_0}(\tau_1)} \right]. \label{e:Tdef}
\end{equation}
(Here the initial distribution in the probability ratio cancels.) 
This functional is the value of a
counting process with antisymmetric increments
\begin{eqnarray}
r^{(2)}_{\sigma',\sigma}(\tau)
&=\ln \left[
\frac{w_{\sigma',\sigma}(\tau)}{w^\text{ad}_{\sigma',\sigma}(\tau)}
\right] \\
&=\ln \left[ \frac{w_{\sigma',\sigma}(\tau) P^*(\sigma;\tau)}{w_{\sigma,\sigma'}(\tau)
  P^*(\sigma';\tau)} \right] \label{e:r2}
\end{eqnarray}
where we have used~\eref{2-7d} to write the adjoint-transformed rate in
terms of the original rate and the quasi-stationary distribution. 

(ii) We can combine both reversal of the protocol and
reversal of the dynamics in the comparison of the probability of a trajectory.
This arises from a counting process defined by the antisymmetric increments
\begin{eqnarray}
r^{(3)}_{\sigma',\sigma}(\tau)
&=\ln \left[
\frac{w_{\sigma',\sigma}(\tau)}{w^\text{ad}_{\sigma,\sigma'}(\tau)}
\right] \\
&=\ln \left[ \frac{P^*(\sigma',\tau))}{P^*(\sigma;\tau)} \right] \label{e:r3}
\end{eqnarray}
and we obtain a functional
\begin{equation} 
\mathcal{Y}(t,\{\sigma\},P_F,P_B)
=\ln \left[
\frac{P^*(\sigma_n;\tau_n) \ldots  P^*(\sigma_{2};\tau_{2})
P^*(\sigma_{1};\tau_{1}) P_F(\sigma_0)}
{P^*(\sigma_{n-1};\tau_{n}) \ldots P^*(\sigma_{1};\tau_{2}) 
P^*(\sigma_{0};\tau_{1}) P_B(\sigma_n)}\right].
\label{e:Ydef}
\end{equation} 
Since $r^{(1)}=r^{(2)}+r^{(3)}$ for all pairs of states we have
$\mathcal{R}_F=\mathcal{T}+\mathcal{Y}$ 
for the forward functional $\mathcal{R}_F$.  Here the rates 
$w_{\sigma',\sigma}(\tau)$ (and corresponding quasi-stationary
probabilities) in $\mathcal{T}$ and $\mathcal{Y}$ are identified with
those of the forward process.

For a physical interpretation of these functionals we follow 
\citeasnoun{Hatano01} and argue that the total heat dissipated can be
divided into two distinct contributions:
\begin{equation}
\mathcal{Q}=\mathcal{Q}_\text{hk}+\mathcal{Q}_\text{ex}.
\end{equation}
Here $\mathcal{Q}_\text{hk}$ is the ``house-keeping heat'' associated
with maintaining a non-equilibrium stationary state for fixed rates
and $\mathcal{Q}_\text{ex}$ is the ``excess heat'' resulting from the
change in rates.  We associate these two contributions with the
quantities measured by the counting processes defined by~\eref{e:r2}
and~\eref{e:r3} so that,
\begin{eqnarray}
\mathcal{T} &=\frac{\mathcal{Q}_\text{hk}}{T} \\
\mathcal{Y} &=\frac{\mathcal{Q}_\text{ex}}{T} + \ln \left[ \frac{
P_F(\sigma_0)}{ P_B(\sigma_n)} \right]
\end{eqnarray}
To support these identifications we remark on two special cases:
\begin{itemize}

\item \emph{Time-independent rates} that do not satisfy the detailed balance
  condition~\eref{2-6}: In this 
  case $\mathcal{Y}$ reduces to
  temporal boundary terms and $\mathcal{R}_F$ and $\mathcal{T}$ measure
  the same quantity (up to boundary terms).

\item Time-dependent rates maintaining \emph{time-dependent detailed
balance} for all times: Now $r^{(2)}(\tau)=0$ for all $\tau$.  Hence
$\mathcal{T}=0$ and $\mathcal{R}_F = \mathcal{Y}$. 

\end{itemize}

We briefly discuss the integral FTs for these functionals. 
The generating function for
$\mathcal{T}=\mathcal{Q}_\text{hk}/T$ is simply given by
\begin{equation}
\langle \,\rme^{-\lambda \mathcal{T}}\, \rangle^F
= \langle \,s \,|\, 
T\{ \rme^{-\int_0^t \tilde{H}_\text{hk}(\lambda,\tau) \, \rmd \tau} \}
\,|\,P_F\, \rangle \label{e:Qhkgen}
\end{equation}
where $\tilde{H}_\text{hk}(\lambda,\tau)$ is the modified Hamiltonian
for the counting process with increments~\eref{e:r2}.  Now
$\tilde{H}_\text{hk}(\lambda,\tau)$ turns out to have the symmetry
property
\begin{equation}
\tilde{H}_\text{hk}(\lambda,\tau) 
= \tilde{H}_\text{hk}^\text{ad}(1-\lambda,\tau) \label{e:Hhksym}
\end{equation}
which allows the simple derivation of a relationship
between the generating functions for forward and adjoint dynamics,
i.e., the analogue of~\eref{e:Rsym} for reversal of dynamics rather
than reversal of protocol.  The corresponding integral
fluctuation relation \cite{Speck05}
\begin{equation}
\langle \,\rme^{- \mathcal{Q}_\text{hk}/T}\, \rangle = 1 \label{e:hkint}
\end{equation}
can be derived almost by inspection from~\eref{e:Qhkgen}
and~\eref{e:Hhksym} and requires no reference to the adjoint process.

Similarly, the generating function for $\mathcal{Y}$ is
\begin{equation}
\langle \,\rme^{-\lambda \mathcal{Y}}\, \rangle_F
= \langle \,s \,| \,
T\{ \rme^{-\int_0^t \tilde{H}_\text{ex}(\lambda,\tau) \, \rmd \tau} \}
\,|\,P_F\, \rangle \label{e:Qexgen}
\end{equation}
where $\tilde{H}_\text{ex}(\lambda,\tau)$ [increments defined
by~\eref{e:r3}] obeys
\begin{equation}
\tilde{H}^T_\text{ex}(\lambda,\tau) =
\tilde{H}_\text{ex}^\text{ad}(\lambda-1,\tau) \label{e:Hevsym}.
\end{equation}
The corresponding integral fluctuation relation reads
\begin{equation}
\langle \,\rme^{- \mathcal{Y}}\, \rangle^F = 1. \label{e:Yint}
\end{equation}
and
with the choice~\eref{e:qs} for
boundary terms (corresponding physically to transitions between steady
states) we obtain \cite{Hatano01}
\begin{equation}
\langle \,e^{-\mathcal{Q}_\text{ex}/T + \Delta \mathcal{S}_\text{sys}}
\,\rangle = 1.
\end{equation}
This ``Hatano-Sasa relation'' can be considered  as a form of
generalized Second Law.  For an experimental realization of 
the corresponding transitions between stationary
states, see \citeasnoun{Trepagnier04}.  The integral relation for ``house-keeping heat''~\eref{e:hkint} was
derived for this situation by \citeasnoun{Speck05} who also provide a
detailed discussion of the interconnections between the three integral
fluctuation theorems.  It may be interesting to further
look into the significance of the full symmetry of the generating
function of the functionals $\mathcal{T}$ and $\mathcal{Y}$.

(iii) A natural further generalization of FTs is the inclusion of other
symmetries in addition to time reversal, see e.g. 
\citeasnoun{Maes06}, \citeasnoun{DeRoeck06}. 
The investigation of the role of internal symmetries
combined with time reversal may provide significant new insights
into the structure of non-equilibrium probability distributions
in specific systems. 

Despite the success story that the exploration of FTs has been,
one should remember that for many processes in nature the 
lack of memory assumed in the dynamical description through Markov 
processes is not an appropriate approximation. However,
since the comparison of probabilities of trajectories (which
lead to the functionals of interest) does not
require Markovian dynamics, the study of time reversal
in non-Markovian dynamics along similar lines is possible and
indicates another very promising direction of research. Some progress 
has been achieved recently within a Langevin approach \cite{Mai06}.

It would
be highly welcome to have more experimental data to probe the
range of validity of FTs as well as their theoretical interpretation
in terms of non-equilibrium thermodynamics.  Of particular interest are
genuine many-particle situations (e.g. dense colloidal systems), 
the description of which cannot be reduced to Markovian effective 
single-particle dynamics.
This would yield significant progress in the understanding of
non-equilibrium phenomena, but also poses a real challenge. To return
to the beginning of this review:
FTs are still mainly a tool to 
quantify in a non-equilibrium setting what in centuries long gone keen 
observers of nature called \emph{horror vacui}. They do not provide a 
recipe to master the horror of waiting for the reader's favourite
example of an unlikely event.

\ack

We would like to thank the many colleagues who have helped develop our
understanding of fluctuation theorems.  In particular, we have enjoyed
fruitful interactions with: Chris Jarzynski, Dragi Karevski, Jorge
Kurchan, Attila R\'akos, Udo Seifert, and Herbert Spohn.  We are
especially indebted to Attila R\'akos for sharing with us the argument
presented in section~\ref{ss:GC} and for detailed comments on the draft
manuscript.  It is also a pleasure to acknowledge the Isaac Newton
Institute, Cambridge for kind hospitality and our fellow-participants
on the programme ``Principles of the Dynamics of Non-Equilibrium
Systems'' for many stimulating discussions.

\appendix

\section[ ]{Toy model: Single-site ASEP}
\label{s:app}

Here we provide a pedagogical example to show how the general
framework established in this review can be applied to a concrete
model.  Specifically, we apply our approach to a simple two-state
model---the single-site asymmetric simple exclusion process (ASEP).  The simplicity
of this model enables us to write down transition matrices and
probability vectors in a compact form and explicitly demonstrate the
emergence of fluctuation symmetries.  The success of recent
experiments on a simple two-state system \cite{Schuler05,Tietz06b}
provides another justification for studying such toy models.

\begin{figure}
\begin{center} 
\psfrag{1}[][]{} 
\psfrag{a}[][]{$\alpha$}
\psfrag{d}[][]{$\delta$} 
\psfrag{bw}[][]{$\beta$}
\psfrag{cw}[][]{$\gamma$} 
\psfrag{I}[][]{} \psfrag{O}[][]{}
\mbox{\subfigure[Site
empty]{\includegraphics*[width=0.21\textwidth]{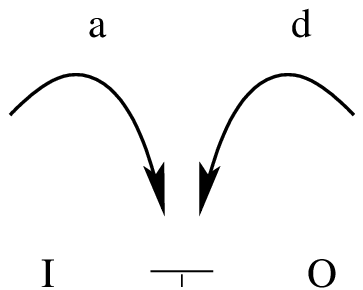}}
\qquad\qquad \psfrag{a}[][]{} \psfrag{d}[][]{} \subfigure[Site
occupied]{\includegraphics*[width=0.21\textwidth]{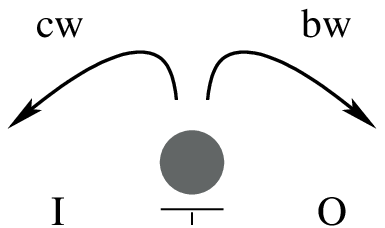}}}
\caption{Schematic representation of a 1 site exclusion process with
two possible states: (a) site empty, a particle can enter from the
left with rate $\alpha$ or from the right with rate $\delta$ (b) site
occupied, particle can exit to the left or right with rates $\gamma$,
$\beta$ respectively.}
\label{f:ASEP1}
\end{center}
\end{figure}

Exclusion processes are paradigmatic lattice-gas models in which each
site can be occupied by at most one
particle \cite{Schmittmann95,Liggett99,Schutz01}, see also the discussion in
section~\ref{sss:state}.  A current of particles can be driven through
the system by imposing asymmetric hopping rates in the bulk and/or at
the boundaries.  Here we consider just a single site which a particle
may enter (subject to the exclusion constraint) or leave in two
directions, as shown in figure~\ref{f:ASEP1}.
Although the stationary state of a two-state model always obeys
detailed balance, the identification of two different mechanisms for
each inter-state transition allows us to think of this as a
non-equilibrium model.  Physically, we have a particle current driven
between two reservoirs of unequal chemical potential.

In the quantum Hamiltonian formalism we define basis vectors
\begin{equation} 
|\,0\,\rangle = \left(
\begin{array}
{c} 1 \\ 0
\end{array} 
\right), 
\qquad 
|\,1\,\rangle = \left(
\begin{array}{c} 0 \\ 
1
\end{array} 
\right),
\end{equation} 
corresponding to the site being occupied by a vacancy or a particle
respectively.  (In a spin analogy one can think of these states as
representing spin-up and spin-down).  For time-independent rates, the
quantum Hamiltonian is given by
\begin{equation} 
H = \left(
\begin{array}{cc} 
\alpha + \delta & -\beta -\gamma \\ 
-\alpha - \delta & \beta + \gamma.
\end{array} 
\right)
\end{equation} 
By inspection one sees that the columns sum to zero as required for a
stochastic matrix.  Only a little more effort yields the steady-state
eigenvector corresponding to the zero eigenvalue:
\begin{equation} 
\left(
\begin{array}{c} 
1-\rho^* \\ 
\rho^*
\end{array} 
\right) 
= \left(
\begin{array}{c} 
\frac{\beta+\gamma}{\alpha+\beta+\gamma+\delta} \\
\frac{\alpha+\delta}{\alpha+\beta+\gamma+\delta}
\end{array} 
\right). \label{e:ss}
\end{equation} 
The mean stationary current of particles from the left-hand reservoir into the site
is given by
\begin{eqnarray} 
\bar{j} &= \alpha(1-\rho^*) - \gamma \rho^* \\ 
&= \frac{\alpha \beta - \gamma \delta}{\alpha+\beta+\gamma+\delta}. \label{e:jstat}
\end{eqnarray} 
Of course this is equal to the mean stationary particle current from the site
into the right-hand reservoir.

It follows from~\eref{e:jstat} that if the rates obey the condition
\begin{equation} 
\alpha \beta = \gamma \delta, \label{e:ASEPeq}
\end{equation} 
the mean stationary particle current is zero.  In this case, the steady state
is in fact an equilibrium state with particle density
\begin{equation}
\rho^\text{eq}=\frac{\alpha}{\alpha+\gamma}=\frac{\delta}{\beta+\delta}.
\end{equation} 
All other choices of rates define a driven non-equilibrium system;
without loss of generality we take $\alpha \beta > \gamma \delta$ so
the mean steady-state particle current flows from left to right.  Of course, even
when~\eref{e:ASEPeq} is obeyed there may be a transient
current flow as the system relaxes from some initial
condition towards equilibrium.  In general, we can also consider
time-dependent rates $\alpha(\tau)$, $\beta(\tau)$, $\gamma(\tau)$,
$\delta(\tau)$ with or without the condition~\eref{e:ASEPeq} at any
$\tau$.

Now consider a \emph{general} current $\mathcal{J}_r$ defined by a counting process as discussed in
section~\ref{ss:counting}.  The modified Hamiltonian which appears in
the generating function [cf.~\eref{2-10}] is given by
\begin{equation}
\tilde{H}_r = \left(
\begin{array}{cc}
\alpha + \delta & -\beta\rme^{-\lambda r_\text{o}} -\gamma \rme^{\lambda r_\text{i}} \\
-\alpha\rme^{-\lambda r_\text{i}} - \delta \rme^{\lambda r_\text{o}}  & \beta + \gamma.
\end{array}
\right)
\end{equation}
Antisymmetry of the current restricts us to two parameters $r_\text{i}$ and
$r_\text{o}$ which determine the specific current to be measured.  In
particular, the current $\mathcal{J}_{\mathcal{Q}}$ (corresponding to the
entropy production of the environment) is defined by the choice
[see~\eref{e:r1}]
\begin{equation}
r_\text{i} = \ln\frac{\alpha}{\gamma}, \qquad
r_\text{o} = \ln\frac{\beta}{\delta}.
\end{equation}
Physically this means that in a given trajectory we weight every
particle hopping move by the ratio of the forward and reverse rates.
It is then straightforward to verify the symmetry relation
\begin{equation}
{\tilde{H}}_\mathcal{Q}^T(\lambda)=\tilde{H}_\mathcal{Q}(1-\lambda). \label{e:HsymA}
\end{equation}
Furthermore it is clear that this relation still holds for
time-dependent rates.  Recalling the analysis of
section~\ref{ss:basic}, we see that~\eref{e:HsymA} ensures the validity
of all transient symmetry relations (see
sections~\ref{ss:int}--\ref{ss:transient}). For example, if we choose
time-dependent rates which satisfy the condition~\eref{e:ASEPeq} for all
times then we can recover the Jarzynski relation~\eref{e:Jar} and the
Crooks' FT~\eref{e:CrooksFT}. On the other hand, if we have
time-independent rates and start the system in the non-equilibrium
stationary distribution given by~\eref{e:ss} we get the
fluctuation theorem for entropy~\eref{e:Sdet}.

In section~\ref{ss:GC} we showed that for time-independent rates and
bounded state-space the asymptotic Gallavotti-Cohen symmetry holds not
just for $\mathcal{J}_{\mathcal{Q}}$ but for other currents too
[cf.~\eref{e:GCFTJ}].  We now give an explicit example of this
symmetry in the context of our single-site ASEP\footnote{See
  \citeasnoun{Derrida04b} for an equivalent discussion for a larger
  exclusion process.} where, of course, the
condition of bounded state space is trivially satisfied.  To be
specific, we consider measuring the integrated input current of
particles from the left-hand reservoir onto the site, denoting this
quantity by $\mathcal{J}_{0}$.  In other words for a given trajectory
we simply count $+1$ when a particle enters from the left and $-1$
when it leaves to the left.  The conjugate field $E_{0}$ is defined
such that applying the reverse field restores equilibrium, i.e., we
require $\alpha e^{-E_{0}/2}$, $\beta$, $\gamma e^{E_{0}/2}$ and
$\delta$ to obey the equilibrium condition
\begin{equation}
\alpha e^{-E_{0}/2} \times \beta = \gamma e^{E_{0}/2} \times \delta.
\end{equation}
This yields
\begin{equation}
E_{0}=\ln\frac{\alpha \beta}{\gamma \delta}
\end{equation}
which physically corresponds to the effective driving field pushing
particles into the system.

The modified Hamiltonian in the generating function for
$\mathcal{J}_{0}$ has $r_\text{i}=1$ and $r_\text{o}=0$:
\begin{equation}
\tilde{H}_{0}(\lambda) = \left(
\begin{array}{cc}
\alpha + \delta  & -\beta -\gamma \rme^{\lambda} \\
-\alpha\rme^{-\lambda} - \delta  & \beta + \gamma
\end{array}
\right).
\end{equation}
The symmetry relation~\eref{e:GCFTH} can then be proved by
simple matrix algebra, as follows.
\begin{eqnarray}
\fl & (P_0^{\text{eq}})^{-1} \tilde{H}_{0}(E_{0}-\lambda) P_0^{\text{eq}}&
\nonumber \\
\fl & \qquad = 
\left(
\begin{array}{cc}
\frac{\beta + \delta}{\beta}  & 0 \\
0 & \frac{\beta + \delta}{\delta}
\end{array}
\right)
\left(
\begin{array}{cc}
\alpha + \delta  & -\beta -\gamma \rme^{(E_{0}-\lambda)} \\
-\alpha\rme^{-(E_{0}-\lambda)} - \delta  & \beta + \gamma
\end{array}
\right)
\left(
\begin{array}{cc}
\frac{\beta }{\beta+ \delta}  & 0 \\
0 & \frac{\delta}{\beta +\delta}
\end{array}
\right) \\
\fl & \qquad = 
\left(
\begin{array}{cc}
\alpha + \delta & -\alpha\rme^{-\lambda} - \delta   \\
-\beta -\gamma \rme^{\lambda}  & \beta + \gamma
\end{array}
\right) \\
\fl & \qquad = \tilde{H}^T_{0}(\lambda)
\end{eqnarray}
A straightforward corollary of this relationship is the eigenvalue
symmetry $\epsilon_{k}(\lambda) = \epsilon_{k}(E_{0}-\lambda)$
which can be explicitly checked on the lowest eigenvalue:
\begin{equation}
\fl
\epsilon_{0}=\frac{\alpha+\beta+\gamma+\delta 
- \sqrt{(\alpha+\beta+\gamma+\delta)^2 + 4[(\beta 
+ \gamma e^\lambda)(\alpha e^{-\lambda} + \delta) 
-(\alpha + \delta)(\beta+\gamma)]}}{2}.
\end{equation}
As detailed in section~\ref{ss:GC} this property of the eigenvalues
leads, for bounded state-space, to the Gallavotti-Cohen symmetry
\begin{equation}
\frac{p(-\mathcal{J}_0)}{p(\mathcal{J}_0)} \sim \rme^{-E_{0} \mathcal{J}_0 }.
\end{equation}
Note that if we had instead considered the current $\mathcal{J}_1$ out
of the system into the right-hand reservoir, we would have obtained a
different equilibrium distribution $P_1^\text{eq}$ for the undriven
system but the same Gallavotti-Cohen symmetry relation (in this case
the conjugate fields are equal, i.e., $E_1=E_0$).

\section*{References}


\end{document}